\documentclass[11pt,a4paper]{article}
\pdfoutput=1
\usepackage[utf8]{inputenc}

\usepackage{amsmath,amssymb}
\usepackage{epsfig,graphicx}
\usepackage{subfigure}
\usepackage{graphicx}
\usepackage{rotating}
\usepackage{cancel}
\usepackage{bm}
\usepackage{color}
\usepackage{comment}
\usepackage{cite}
\usepackage{psfrag}
\usepackage[pagebackref]{hyperref}
\usepackage[normalem]{ulem}

\newcommand{\diff}{\mathrm{d}}

\newcommand{\me}{\mathrm{e}}

\renewcommand\[{\left[}

\usepackage{textpos}

\begin{document}
\numberwithin{equation}{section}
\title{
\vspace{2.5cm} 
\Huge{\textbf{
Tracking axion-like particles\\ at the LHC
\vspace{0.5cm}}}}

\author{Gonzalo Alonso-\'Alvarez$^{1}$, Joerg Jaeckel$^{2}$, and Diego D. Lopes$^{3}$\\[2ex]
\small{\em $^1$McGill University Department of Physics \& McGill Space Institute,}\\ \small{\em 3600 Rue University, Montr\'eal, QC, H3A 2T8, Canada}\\[0.5ex]
\small{\em $^2$Institut f\"ur theoretische Physik, Universit\"at Heidelberg,} \\
\small{\em Philosophenweg 16, 69120 Heidelberg, Germany}\\[0.5ex]
\small{\em $^3$Centro de Ci\^encias Naturais e Humanas,} \\
\small{\em Universidade Federal do ABC, Santo Andr\'e-SP, Brasil}\\[0.8ex]
}

\date{}
\maketitle

\begin{abstract}
\noindent
Highly boosted axion-like particles decaying into photon pairs are notoriously hard to detect at the LHC. 
The collimated decay photons cannot be individually reconstructed using only electromagnetic calorimeter information, making the signal less distinguishable from background.
In this note we propose a search strategy to address this issue, exploiting the fact that a fraction of the decay photons convert into electron-positron pairs inside the tracking detector. 
The resulting tracks can be resolved with high resolution, allowing to separate the two collimated photons and resolve a displaced decay vertex. 
To demonstrate the effectiveness of this approach, we apply it to ALPs in the challenging MeV-GeV range produced via vector boson fusion. We find that such a search could give sensitivity to untested parameter space.
\end{abstract}

\newpage
\tableofcontents

\section{Introduction}
The $\sim 10\, {\rm MeV}-10\,{\rm GeV}$ mass region is one of the  most challenging targets to search for feebly interacting particles (FIPs)~\cite{Agrawal:2021dbo}. 
Somewhat surprisingly, relatively strong couplings are most problematic to test. This is due to a combination of several factors.
The significant mass of the prospective particle requires reasonably high energy experiments, suggesting accelerator-based fixed-target or collider experiments. 
The former, however, suffer from the relatively short lifetimes of the FIPs causing the visible decay products to be absorbed in the shielding, thereby suppressing the signal. 
Moreover, they usually do not have enough energy to go far beyond the GeV scale.
At present, low energy/high intensity collider $e^{+}e^{-}$ experiments such as Belle II~\cite{Dolan:2017osp,Belle-II:2020jti} seem most promising in this direction.

High energy colliders such as the LHC, on the other hand, have plenty of energy and sufficient luminosity. 
Indeed, for an axion-like particle (ALP) with a mass of $200\,{\rm MeV}$ and a two-photon coupling $g_{a\gamma\gamma}=1\,{\rm TeV}^{-1}$, the production cross section is $\sim pb$ such that a ${\rm few}\times 10^5$ ALPs would have already been produced up to run 2.
This is clearly not a small number.
However, here the problem is that the typical signal is hard to identify amidst the background. A prime example for this are ALPs coupled to two photons. In principle their two photon decay should allow to perform a search for a peak in the invariant mass spectrum, similar to what was done for the Higgs discovery~\cite{ATLAS:2012yve,CMS:2012qbp}.
However, for masses in the $\sim 10\, {\rm MeV}-10\,{\rm GeV}$ range, ALPs are typically produced with a relatively high boost. As a consequence, the two decay photons are highly collimated, making their individual identification particularly difficult. 
At the lower end of this mass region, the photons become so collimated that they effectively look like a single photon originating from the point where the ALP was produced.
This clearly prevents the determination of an invariant mass but also makes it difficult to identify a (possibly only slightly) displaced vertex. 
On top of this, the $\sim 100\,{\rm MeV}-{\rm GeV}$ region is riddled with backgrounds from mesons that have similar or identical decay products.
It is therefore imperative to find strategies to mitigate such backgrounds.\footnote{One possibility is to select special production channels. For example, if the two-photon coupling is inherited from a generic combination of electroweak  couplings, ALPs may be efficiently produced in $Z$ decays, which can then be identified~\cite{Jaeckel:2015jla}.}

Work in this direction has already achieved some successes in the past.
ATLAS~\cite{ATLAS:2019dpa} and CMS~\cite{CMS:2020cmk} have implemented dedicated triggers geared towards improving their sensitivity to low-mass diphoton resonances.
In this line, Ref.~\cite{Knapen:2021elo} has proposed modifying the currently used photon isolation requirements at the trigger level to avoid discarding potential ALP decay events containing collimated photon pairs. More generally, other studies have proposed using modified calorimetric shower shape variables tailored to collinear multi-photon or \emph{photon-jet} objects~\cite{Ellis:2012zp,Toro:2012sv,Allanach:2017qbs,Sheff:2020jyw,Wang:2021uyb,Ren:2021prq}.

Collimated diphotons  can also be discriminated by their increased likelihood of producing a converted photon event, i.e. an increase probability that a photon converts into an electron-positron pair.
This was proposed in~\cite{Dasgupta:2016wxw}, originally to distinguish diphoton and 4-photon final states, thereby detecting the presence of intermediate (pseudo)scalars in the decay of a heavier resonance.

As in~\cite{Dasgupta:2016wxw}, in this work we make use of photon conversions but aim to develop a methodology that can be used to look for a light resonance largely independently of its production channel. 
Going beyond counting the number of conversion events, we propose using the exquisite spatial resolution of the ATLAS tracking detector to identify collimated photon pairs originating from the \emph{displaced} decay of a new (pseudo)scalar state.
For that, we exploit the fact that a sizeable fraction of photons convert into $e^{+}e^{-}$-pairs inside the tracker and thus leave a visible track in it.
The superior resolution of the tracker compared to the electromagnetic calorimeter allows to separate the tracks of two very collimated photons that effectively behave as a single photon in the calorimeter.
To eliminate contamination from SM hadronic decays, we require the decay of the new physics resonance to be displaced from the original interaction vertex.
Once more, the fine resolution of the tracker facilitates the task of identifying such a displaced vertex.

To showcase its potential, we apply our search strategy to a model of axion-like particles produced through photon and vector boson fusion and decaying to photon pairs. We find that already existing data provides sensitivity to a significant range of so-far untested parameter space. 
While for concreteness we have focused on a specific example particle and production mechanism, we stress that the same strategy is applicable to ALPs produced through other mechanisms as well as other particles featuring a delayed decay into two or more photons.
Moreover, while we focus on ATLAS as a concrete example for the detector, the same technique should also be applicable to CMS or other instruments with the appropriate adjustments.

This paper is structured as follows.
We start by specifying the benchmark ALP model and characterize the ALP production and decay processes at the LHC in Sec.~\ref{sec:model}.
Our proposed search strategy is described in detail in Sects.~\ref{sec:conversiongen} and \ref{sec:conversionimp}.
Sec.~\ref{sec:results} is devoted to the discussion of its application to the ALP model under study.
Conclusions are drawn in Sect.~\ref{sec:conclusions}.

\section{Physics case study: axion-like particles at the LHC}\label{sec:model}

\subsection{Benchmark ALP model}
The benchmark model (cf., e.g.,~\cite{Agrawal:2021dbo}) for many ALP studies is that of an ALP coupled electromagnetically to the SM via the Lagrangian
\begin{equation}\label{eq:Lag_photons}
    {\mathcal{L}}=\frac{1}{2}\left(\partial_{\mu}a\right)^2-\frac{1}{2}m^{2}_{a}-\frac{1}{4}g_{a\gamma\gamma} \, a \, F_{\mu\nu} \tilde{F}^{\mu\nu}.
\end{equation}
For high-energy probes, as the ones of interest here, couplings to other electroweak gauge bosons can also play an important role.
Thus, it is necessary to specify those in a consistent manner.
For concreteness, we consider an ALP coupled to the hypercharge gauge boson field strength $B^{\mu\nu}$ (see, e.g.,~\cite{Jaeckel:2015jla,Bauer:2017ris,Alonso-Alvarez:2018irt}), 
\begin{align}
\label{eq:benchmark}
    {\mathcal{L}} &\supset -\frac{1}{4} \frac{1}{c_W^2} g_{a\gamma\gamma} \, a \, B_{\mu\nu} \tilde{B}^{\mu\nu} \nonumber \\
    &= -\frac{1}{4}g_{a\gamma\gamma} \, a \, F_{\mu\nu} \tilde{F}^{\mu\nu} -\frac{1}{4} \frac{s_W^2}{c_W^2} g_{a\gamma\gamma} \, a \, Z_{\mu\nu} \tilde{Z}^{\mu\nu} +\frac{1}{4} \frac{2s_W}{c_W} g_{a\gamma\gamma} \, a \, F_{\mu\nu} \tilde{Z}^{\mu\nu}.
\end{align}
The second line shows the couplings to the photon and the $Z$ boson after electroweak symmetry breaking (EWSB).
In the Lagrangian above, $s_W$ and $c_W$ represent the sine and cosine of the Weinberg angle and the coupling strengths are normalized to recover the same photon coupling as in Eq.~\eqref{eq:Lag_photons}.

\begin{figure}[t]
\centering
\includegraphics[width=0.7\textwidth]{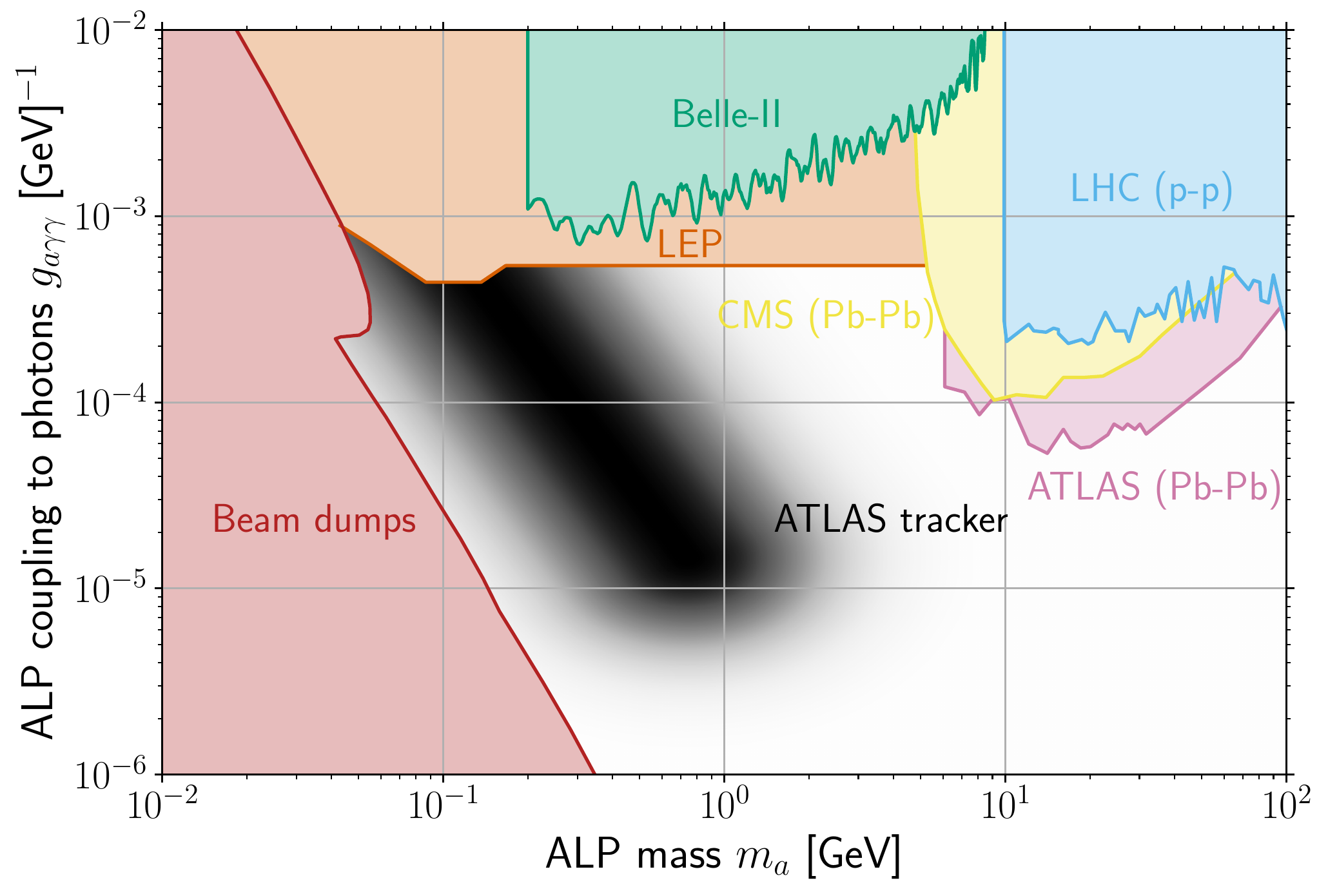}
\caption{Existing constraints on electromagnetically coupled ALPs in the relevant $\sim {\rm MeV}- {\rm multi\,\, GeV}$ mass range.
We show bounds from beam dump searches~\cite{Bjorken:1988as,Blumlein:1990ay,NA64:2020qwq}, LEP~\cite{Jaeckel:2015jla}, Belle-II~\cite{Belle-II:2020jti}, Pb-Pb collisions at CMS~\cite{CMS:2018erd} and ATLAS~\cite{ATLAS:2020hii}, and p-p collisions at the LHC~\cite{Knapen:2016moh}.
The black shaded region represents the region that we target in our study, based on a simplistic estimation using the ALP decay length and production cross section at the LHC.}
\label{fig:parameter_space_sketch}
\end{figure}

The mass range of interest to us is targeted by an extensive program of searches at high-energy experiments looking for axion-like particles with electromagnetic couplings.
The impact of these searches in the $m_a$ vs.~$g_{a\gamma\gamma}$ is shown in Fig.~\ref{fig:parameter_space_sketch}.
At the lowest masses considered here, $\sim(10-100)\,\mathrm{MeV}$, ALPs can be copiously produced at electron and proton fixed-target experiments~\cite{Dobrich:2015jyk}.
A combination of experiments~\cite{Riordan:1987aw,CHARM:1985anb,Bjorken:1988as,Blumlein:1990ay,NA64:2020qwq}, among which NA64~\cite{NA64:2020qwq}, NuCal~\cite{Blumlein:1990ay}, and E137~\cite{Bjorken:1988as} give the strongest limits, shape the red region labelled as ``Beam dumps'' in Fig.~\ref{fig:parameter_space_sketch}.
At intermediate masses, $\sim(0.1-{\rm few})\,\mathrm{GeV}$, $e^+e^-$ colliders take over as the prime environment to look for ALPs.
The orange region shows the LEP~\cite{L3:1994shn,OPAL:2002vhf} limits derived in~\cite{Jaeckel:2015jla,Knapen:2016moh}, while the green region corresponds to the exclusion by Belle-II~\cite{Dolan:2017osp,Belle-II:2020jti}.
By now superseded bounds from PrimEx~\cite{Aloni:2019ruo}, CLEO~\cite{CLEO:1994hzy}, Babar~\cite{BaBar:2010eww}, and CDF~\cite{CDF:2013lma} are not shown in Fig.~\ref{fig:parameter_space_sketch} to avoid clutter.
The most massive ALPs, $\gtrsim {\rm few}\,{\rm GeV}$, are best searched for at the LHC~\cite{dEnterria:2021ljz}.
The enhanced $\gamma\gamma$ cross-section in ultra-peripheral heavy-ion (Pb-Pb) collisions~\cite{Knapen:2016moh,Knapen:2017ebd} allows CMS~\cite{CMS:2018erd} and ATLAS~\cite{ATLAS:2020hii} to place stringent constraints above a few GeV.
At the highest masses, searches focus on ALPs produced via photon or vector boson fusion in p-p collisions, which subsequently decay to photon pairs.
A combination of multiple searches~\cite{CMS:2011uvc,CMS:2011bsw,ATLAS:2011ab,CMS:2012cve,ATLAS:2012yve,ATLAS:2012fgo,ATLAS:2014jdv,ATLAS:2017ayi,CMS:2020rzi} and reinterpretations~\cite{Jaeckel:2012yz,Jaeckel:2015jla,Knapen:2016moh,Knapen:2017ebd,Bauer:2017ris,Bauer:2018uxu,Florez:2021zoo} allow to test ALPs with masses up to above the TeV scale.

In the present study, we focus on the model of an electromagnetically coupled ALP with Lagrangian Eq.~\eqref{eq:benchmark}.
That said, our proposed methods are applicable in a variety of other situations where two or more highly collimated photons are the primary signal. 
Examples include more general axion-like particles with additional couplings, which can be produced by different processes such as gluon fusion, decays of Higgs bosons or mesons, etc.
Our strategy can thus help to adapt other existing searches for ALPs~\cite{Mimasu:2014nea,Mariotti:2017vtv,Alonso-Alvarez:2018irt,CidVidal:2018blh,Gavela:2019cmq} to be effective in the $10\,\mathrm{MeV}-10\,\mathrm{GeV}$ region.

\subsection{ALP production and decay}
\begin{figure}[t]
\centering
\includegraphics[width=0.45\textwidth]{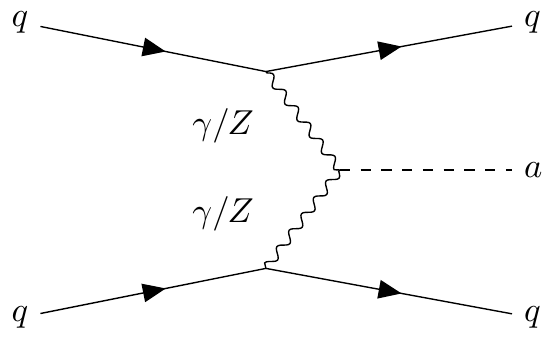} 
\caption{Vector boson fusion process by which an electromagnetically coupled ALP is produced in proton-proton collisions at the LHC.}
\label{fig:VBF_diagram}
\end{figure}

The leading process by which electromagnetically coupled ALPs can be produced at the LHC is photon and vector boson fusion.
The diagrams corresponding to these processes are shown in Fig.~\ref{fig:VBF_diagram}.
In our benchmark model where the ALP couples to the hypercharge gauge boson, possible channels are $\gamma\gamma \rightarrow a$, $ZZ \rightarrow a$, and $\gamma Z \rightarrow a$.
At high energies, the three channels contribute at a similar level.

In order to characterize the kinematic properties of the produced ALPs, we run a simulation using \texttt{MadGraph5}~\cite{Alwall:2014hca}. The process $p p \to a (\to \gamma \gamma) j j$ was simulated at $\sqrt{s} = 13$ TeV with a transverse momentum requirement of $10$~GeV for each outgoing photon. 
No separation requirement was imposed on the photon pairs at this stage.
The production cross section found,
\begin{equation}
    \sigma(pp\xrightarrow{VBF} a) \simeq 18\,\mathrm{fb}\,\left(\frac{g_{a\gamma\gamma}}{0.1\,\mathrm{TeV}^{-1}}\right)^2,
\end{equation}
is largely independent of the ALP mass in the range of interest.
This value can be used to estimate a lower limit to the values of the ALP-photon coupling accessible at the LHC.
A coupling of $g_{a\gamma\gamma}\sim 0.01\,\mathrm{TeV}^{-1}$ results in $50-60$ ALPs being produced through vector boson fusion at the Run 3 of the LHC. 
We use this figure as a guideline to estimate the target region of our search strategy in the ALP parameter space shown if Fig.~\ref{fig:parameter_space_sketch}, although the actual sensitivity depends on the precise search and detector efficiencies, as discussed below.

\begin{figure}[t]
\centering
\includegraphics[width=0.6\textwidth]{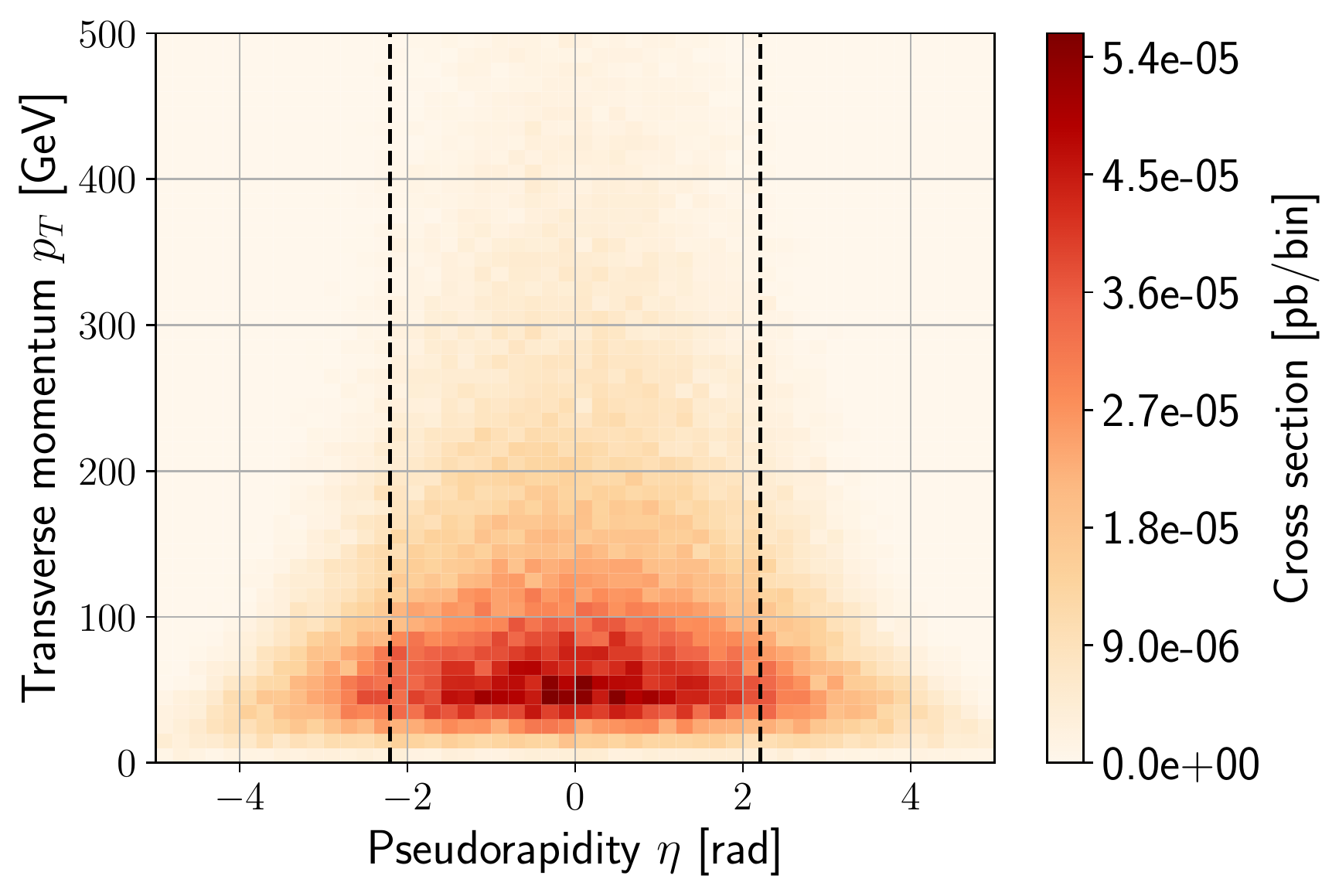} 
\caption{Pseudorapidity and transverse momentum distribution of ALPs produced through VBF from $pp$ collisions at a center of mass energy of 13 TeV.
The vertical dashed lines show the angular extent of the ATLAS TRT coverage.
For this figure we fix the mass of the ALP to be $m_a=100$~MeV (the distribution does not appreciably change for values of $m_a$ between $10$~MeV and $10$~GeV).}
\label{fig:pt_eta_distribution}
\end{figure}

As we will see, the most important kinematic variables for our analysis are the ALP transverse momentum ($p_T$) and pseudorapidity ($\eta$).
The combined distribution for these quantities is shown in Figure~\ref{fig:pt_eta_distribution}.
The production is predominantly central, with around 80\% of the events being contained in the $-2.2\leq \eta \leq 2.2$ region which corresponds to the TRT coverage.
The $p_T$ distribution is mostly concentrated at low momenta $p_T\lesssim 100$~GeV.
Yet around 25\% of the produced ALPs have a transverse momentum larger than $150$~GeV (the single-photon trigger at ATLAS was set to $140$~GeV during Run 2~\cite{ATLAS:2019dpa}).

After the ALPs are produced, they decay almost exclusively into two photons (loop-induced decays\footnote{See~\cite{Srednicki:1985xd,Chala:2020wvs,Bonilla:2021ufe} for calculations of the loop-induced coupling and its renormalization group running.} to charged fermion/antifermion pairs have small rates and we neglect them).
Decays to other gauge bosons would be possible for ALPs with $m_a\gtrsim 100$~GeV, but in this work we are interested in lighter ALPs with masses below $10$~GeV.
The decay rate of an ALP to two photons is given by (cf., e.g.,~\cite{Jaeckel:2015jla,Bauer:2017ris,Alonso-Alvarez:2018irt})
\begin{equation}\label{eq:Gamma_ALP}
    \Gamma(a\rightarrow \gamma\gamma) = \frac{g_{a\gamma\gamma}^2 m_a^3}{64\pi}.
\end{equation}
The inverse of this quantity gives the ALP lifetime, or equivalently the decay length in the relativistic limit.
For the values of ALP mass and photon coupling of interest here, the ALP can have a macroscopic decay length, especially when the boost factor $p_a / m_a$ is taken into account.
For $p_a = 150$~GeV, which is representative of the typical momentum of ALPs in our analysis, the decay length is
\begin{equation}\label{eq:l_decay}
    \ell_{\rm decay} \simeq 40\,\mathrm{cm}\,\left( \frac{p_a}{150\,\mathrm{GeV}} \right) \left( \frac{0.2\,\mathrm{GeV}}{m_a} \right)^4 \left( \frac{0.1\,\mathrm{TeV}^{-1}}{g_{a\gamma\gamma}} \right)^2.
\end{equation}
Decay lengths between a centimeter and a few meters can result in ALP decay vertices that can be identified as displaced in the ATLAS tracker.
ALPs with longer decay lengths typically decay outside of the inner detector and thus cannot leave any track in the tracker.
A lower limit on the decay length is necessary in order to eliminate background, as we will see.
This requirement, together with a sufficient production cross section, delineates a rough region in the ALP parameter space where our proposed search strategy can be applicable, this is indicated as the black shaded region in Fig.~\ref{fig:parameter_space_sketch}.

\section{Identifying highly collimated photon pairs and displaced vertices with the ATLAS tracker -- general strategy outline}\label{sec:conversiongen}

The main challenge faced by low-mass ALP searches at the LHC is to distinguish signal events with two closely collimated photons from single photon events. 
Our strategy, described in detail below, exploits the fact that the trajectory of photons that convert inside the tracker into $e^+e^-$ pairs can be determined with the higher resolution of the tracker compared to the electromagnetic calorimeter. 
Especially when both photons convert, this not only allows to more readily distinguish two-photon and single-photon events, but also to resolve the ALP decay vertex more precisely.
This makes it possible to identify the ALP decay as displaced and thus to significantly reduce the background affecting our search.

\subsection{The collimation challenge}
\begin{figure}[t]
\centering
\includegraphics[width=0.6\textwidth]{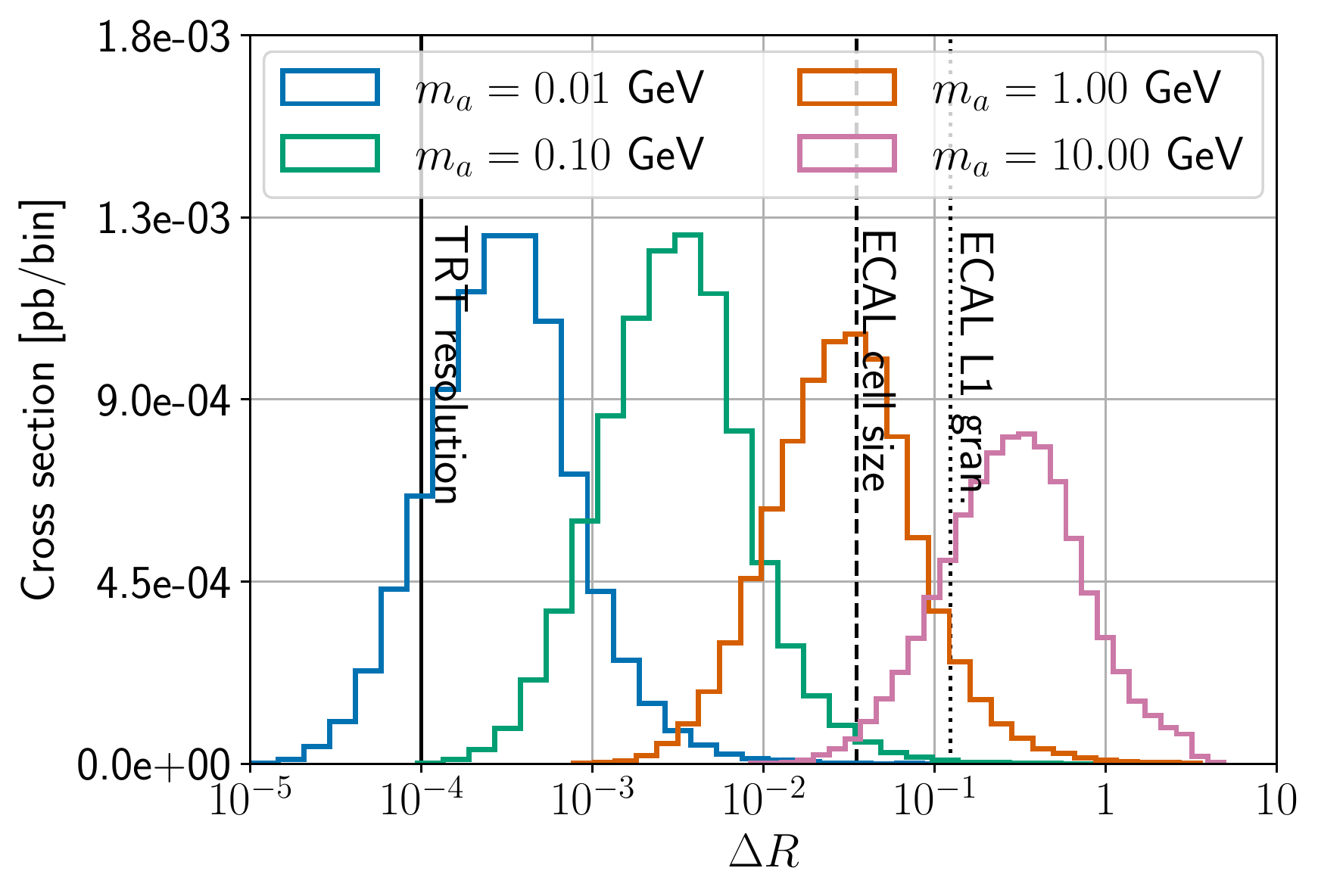} 
\caption{$\Delta R$ distribution of photon pairs produced in the decays of ALPs generated via VBF at the LHC, for a series of ALP masses and fixed $g_{a\gamma\gamma}=0.1\,\mathrm{TeV}^{-1}$.
The vertical solid line corresponds to the angular resolution of the ATLAS TRT.
The vertical dashed line shows the angular resolution of a ECAL tower used for photon selection at the L1 trigger.
The vertical dotted line represents the granularity of a single ECAL cell in the middle layer.
ALPs with $m_a\lesssim 1$~GeV decay into photons which are usually too collimated to pass isolation requirements at the ATLAS ECAL.
Note that in our simulations a cut of $p_{\rm T}\geq 10$~GeV is placed on the individual photon momenta, but in this figure there is no additional cut on the combined $p_{\rm T}$.
}
\label{fig:DeltaR_distribution}
\end{figure}

ALPs with sub-GeV masses are produced at the LHC with a large boost.
This, together with their relatively low mass, makes their decay products highly collimated.
For the decay channel at hand, $a\rightarrow\gamma\gamma$, the angular separation between the two photons can be estimated to be
\begin{equation}
    \Delta R \sim \frac{2m_a}{p_a}
\end{equation}
in the small angle limit.
The quantity $\Delta R = \sqrt{\Delta\eta^2 + \Delta\phi^2}$ is commonly used in collider experiments as a measure of the angular separation between two particles when $\Delta\eta$ and $\Delta\phi$ are both small.
If $p_a\sim\mathcal{O}(100\,\mathrm{GeV})$, the angular separation is thus expected to fall below the $10^{-2}$ level for ALPs with masses below around $1$~GeV.
The actual $\Delta R$ distribution for our simulated events shown in Figure~\ref{fig:DeltaR_distribution} follows this trend.

The high collimation of the photon pairs poses a formidable challenge to LHC searches for ALPs lighter than $\sim 1$~GeV.
As detailed in~\cite{ATLAS:2016ecu,ATLAS:2018fzd,ATLAS:2019dpa}, the standard photon reconstruction and identification at the Run-2 ATLAS phase~\cite{ATLAS:2019dpa} mainly uses information recorded by the ECAL, whose middle layer is made up of cells of size $\delta\eta\times\delta\phi = 0.025\times0.0245$~\cite{ATLAS:2016ecu}, or $\Delta R \simeq 0.035$.
At the L1 trigger and reconstruction levels though, the actual granularity of the ECAL used in the EM cluster selection is larger, $\Delta \eta \times \Delta \phi = 0.075 \times 0.123$~\cite{ATLAS:2016ecu}, corresponding to $\Delta R \simeq 0.14$.
In Figure~\ref{fig:DeltaR_distribution}, the vertical dashed and dotted lines allow to compare these values with the separation of the photon pairs in ALP decay.
It is clear that ALPs with masses below $1$~GeV result in photon pairs that are too collimated to be identified as two separated photons.
In most cases, both photons will end up in the same ECAL cell, making their energy deposition essentially indistinguishable from that of a single photon with the combined energy of the two.

\subsection{The tracker advantage}
Photons can convert into electron-positron pairs in the electromagnetic fields of atoms inside materials.
For high-energy photons this conversion occurs without any significant energy and/or momentum transfer\footnote{At sufficiently high energies the momentum transfer required because the electron and positron are massive is rather small.}.
Therefore the $e^+e^-$ pair initially
carries the energy and momentum of the incoming photon.
The tracker detectors at ATLAS (and CMS) make use of this conversion to determine the trajectory of converted photon, as electrons and positrons, being charged particles, leave tracks in a suitable detector component~\cite{ATLAS:2018fzd}. 
At ATLAS, the probability for a photon to convert inside the tracker is sizable.
As we will see in Sec.~\ref{sec:conversion}, depending on the angle, up to $\sim 50\%$ of photons convert before they leave the tracker~\cite{ATLAS:2018fzd}.
It is thus likely that both photons from an ALP decay convert into electron-positron pairs that leave tracks in the tracker.

The ATLAS Inner Detector consists of the Pixel and SemiConductor Tracker (SCT) detectors and the Transition Radiation Tracker (TRT).
Its main purpose is to reconstruct the tracks of charged particles that travel through it.
It features an incredibly high position measurement accuracy, which in the TRT can reach $0.1$~mm~\cite{ATLAS:2017jwu,2017NIMPA.845..257M}.
The barrel of the TRT extends to $|z|<720$~mm has an inner and outer radius of $560$ and $1080$~mm, respectively.
The two endcaps cover the range $827<|z|<2774$~mm and $617<r<1106$~mm~\cite{2017NIMPA.845..257M}.
This means that the angular resolution of the TRT is around
\begin{equation}\label{eq:TRT_DeltaR}
    \Delta R \sim \frac{0.1\,\mathrm{mm}}{1000\,\mathrm{mm}}\sim 10^{-4},
\end{equation}
This can be compared with the pixel size of the ECAL, which is smallest in the second layer of the LAr, with $\Delta\eta\times\Delta\phi = 0.025\times0.0245$~\cite{ATLAS:2016ecu} (the strips in the first layer have a finer $\eta$ resolution but worse $\phi$ one, $\Delta\eta\times\Delta\phi = 0.0031\times0.098$).
The ATLAS TRT thus offers great prospects for the separation of highly-collimated photons.
As can be seen in Fig.~\ref{fig:DeltaR_distribution}, this angular resolution can help separate photon pairs from ALP decays for ALPs with masses as low as $0.01$~GeV.

The use of the tracker in addition to the ECAL offers great perspectives for the separation of highly collimated photons coming from the decay of a new resonance.
That said, such a new physics signal can be mimicked by many SM processes leading to two photons, such as decays of $\pi^0$, $K^0$, and other neutral hadrons that are copiously produced at the LHC. 
It is therefore necessary to identify a strategy that can help distinguish the photons coming from ALP decays from those generated by hadronic decays. 

\subsection{Locating displaced vertices with the tracker}
Locating a displaced vertex is the most useful way to suppress the backgrounds that affect our search.
The most pressing background is the one caused by neutral pions, which are plentiful at the LHC and have a significant branching fraction into two photons.
Fortunately, their short lifetime produces a decay that is prompt even at LHC energies. 
Thus, identifying a displaced vertex as the origin of the photon pair allows to reduce this background and that from other short-lived hadronic resonances.
The main remaining background is that due to long-lived kaons ($K_{L}$) which, as we discuss in Sec.~\ref{sec:irreducible_backgrounds}, is sufficiently small after all the kinematic cuts of our search are applied.

Beyond separating the two photons from each other, the greater resolution of the tracker also allows for a precise vertex location.
The presence of the track left by the electron-positron pair allows to reconstruct the ``trajectory'' of the photon from the point of conversion.
The determination of the intersection point of the trajectories of the two photons can then be used to decide whether the decay is displaced. 

For small opening angles $\delta\phi$ between the two photons, a very naive parametric estimate for the precision with which the decay vertex can be located is given by
\begin{equation}
\Delta \sim \frac{r_{\rm track}}{\delta\phi}\sim r_{\rm track}\,\frac{p_{a}}{m_{a}},
\end{equation}
where $r_{\rm track}$ stands for the accuracy in the determination of the track and $p_a$ and $m_a$ are the ALP momentum and mass, respectively.
This makes it explicit that at high ALP momenta and small ALP masses, the exquisite resolution of the tracker is required to make any confident statements about the displaced nature of a decay.

Another possible way to reject background is to reconstruct the invariant mass of the photon pair, which allows to discriminate the known hadronic resonances.
However, this is a difficult task to carry out for very collimated photons and evaluating its feasibility requires a careful study.
We do not pursue such a study in this work as our estimates show that the displaced vertex strategy achieves a sufficient background rejection at the level of our analysis, but leave this possibility for future work.

\section{Identifying highly collimated photon pairs and displaced vertices with the ATLAS tracker -- implementation}\label{sec:conversionimp}

In what follows we develop the ideas presented in the previous section to identify highly collimated photon pairs and displaced vertices with the ATLAS tracker, and show how this strategy can be a very powerful tool to search for ALPs at the LHC.

\subsection{Photon triggers, identification, and isolation criteria}

As already discussed, we propose exploiting the tracking information recorded by the ATLAS inner detector, with an angular resolution as low as $\sim 10^{-4}$ (see Eq.~\eqref{eq:TRT_DeltaR}), to distinguish collimated photon pairs.
Our proposal entails a non-trivial modification of the photon identification and isolation pipeline at ATLAS, which is described in~\cite{ATLAS:2016ecu,ATLAS:2018fzd,ATLAS:2019dpa}. We highlight the main points in the analysis that need to be adapted to enable our search strategy.

A few basic considerations have to be made towards putting this idea into practice.
The first point to asses is whether trigger requirements allow for diphoton events under study to be recorded.
In most cases, the photon pair is collimated enough to fall within a single trigger tower and the standard L1 (hardware) trigger on isolated photons will be satisfied by the diphoton event.

The high-level (software) trigger further performs photon identification and isolation tests as detailed in~\cite{ATLAS:2016ecu,ATLAS:2018fzd,ATLAS:2019dpa}.
Loose identification criteria are based on shower shape development in the middle ECAL layer, which would reject diphoton events that are not collimated enough to fall within a single calorimeter cell~\cite{ATLAS:2019dpa}.
That said, as we will see, the collimation of the majority of the events of interest for the ALP model is enough for both photons to fall within a single ECAL cell.
In fact, most of the events have an angular separation below $\Delta\eta=0.0031$, which is the size of the strips in the first ECAL layer that are used for tight identification criteria.

Furthermore, a set of isolation requirements are placed on photon candidates~\cite{ATLAS:2009zsq,ATLAS:2019qmc}.
These typically place a limit on the amount of extra energy in the angular vicinity of the photon candidate.
Given that the signal photons arise from the decay of a singly-produced particle, we do not expect isolation requirements to result in any significant signal event rejection.
This is unlike the potential backgrounds, which mostly originate from hadronic states that are more likely to originate within a QCD jet.

One may worry that the presence of two collimated photons in our signal can lead to the event not passing the isolation requirements.
We do not expect that to be the case given the high collimation of the photon pair.
To gauge any potential negative impact, we place a cut on the angular separation between the photons and confirm that the majority of our signal events pass the most restrictive requirement of $\Delta\eta\leq 0.0031$, corresponding to the finest angular resolution of the ECAL.
Though this assures that calorimetric isolation variables do not negatively affect our signal, it is worth noting that tracking information is also employed at the photon isolation stage.
Thus, a more detailed experiment-level analysis of the impact of tracking isolation variables in our search should be carried in the future.

Beyond the reconstruction and identification triggers described above and in order to manage the huge luminosity, ATLAS currently places a $140$~GeV cut at the trigger level on the transverse momentum ($p_T$) of isolated single-photon events~\cite{ATLAS:2019dpa}.
In our case, this trigger requirement has to be satisfied by the combined momenta of the photon pair or, equivalently, by that of the decaying ALP.
As can be calculated from the distribution shown in Fig.~\ref{fig:pt_eta_distribution}, this shrinks the total signal cross section to about $25\%$ of its original size, but at the same time provides a crucial background rejection.
As will be shown in Sec.~\ref{sec:irreducible_backgrounds}, a $p_T$ cut in the $100-150$~GeV range together with a displacement requirement on the ALP decay vertex are enough to render the search background free to the best of our knowledge.

We also note that, during reconstruction (cf.~\cite{ATLAS:2016ecu,ATLAS:2018fzd,ATLAS:2019dpa}), after a candidate photon event is identified based on its energy deposition in the ECAL, inner detector tracks and conversion vertices are matched to the ECAL cluster.
Currently, only one matching candidate conversion vertex is kept for each ECAL cluster, which precludes the possibility of distinguishing between single and multiple photon events.
Allowing for more than one conversion vertex to be linked to a single ECAL cluster would allow to identify collimated photon pair candidates (cf.~\cite{Dasgupta:2016wxw}). 
If this could be done at the trigger level, it is conceivable that the aforementioned $p_T$ threshold could be significantly lowered.

\subsection{Separation of converted photons}

Once the events are triggered on and thus recorded, we have access to their full tracking information which can be used to discern whether the track matches that left by a single or a pair of photons.
In practice, photons only leave a track in the TRT after they convert into an electron/positron.
We will discuss this conversion probability and its dependence on the photon momentum and direction in Sec.~\ref{sec:conversion}.
For now, we focus on the inner detector's ability to distinguish two separate tracks on an $a\rightarrow\gamma\gamma$ decay where both photons convert.

\begin{figure}[t]
\centering
\renewcommand{\arraystretch}{1}
\begin{tabular}[t]{cc}
	\includegraphics[width=0.49\textwidth]{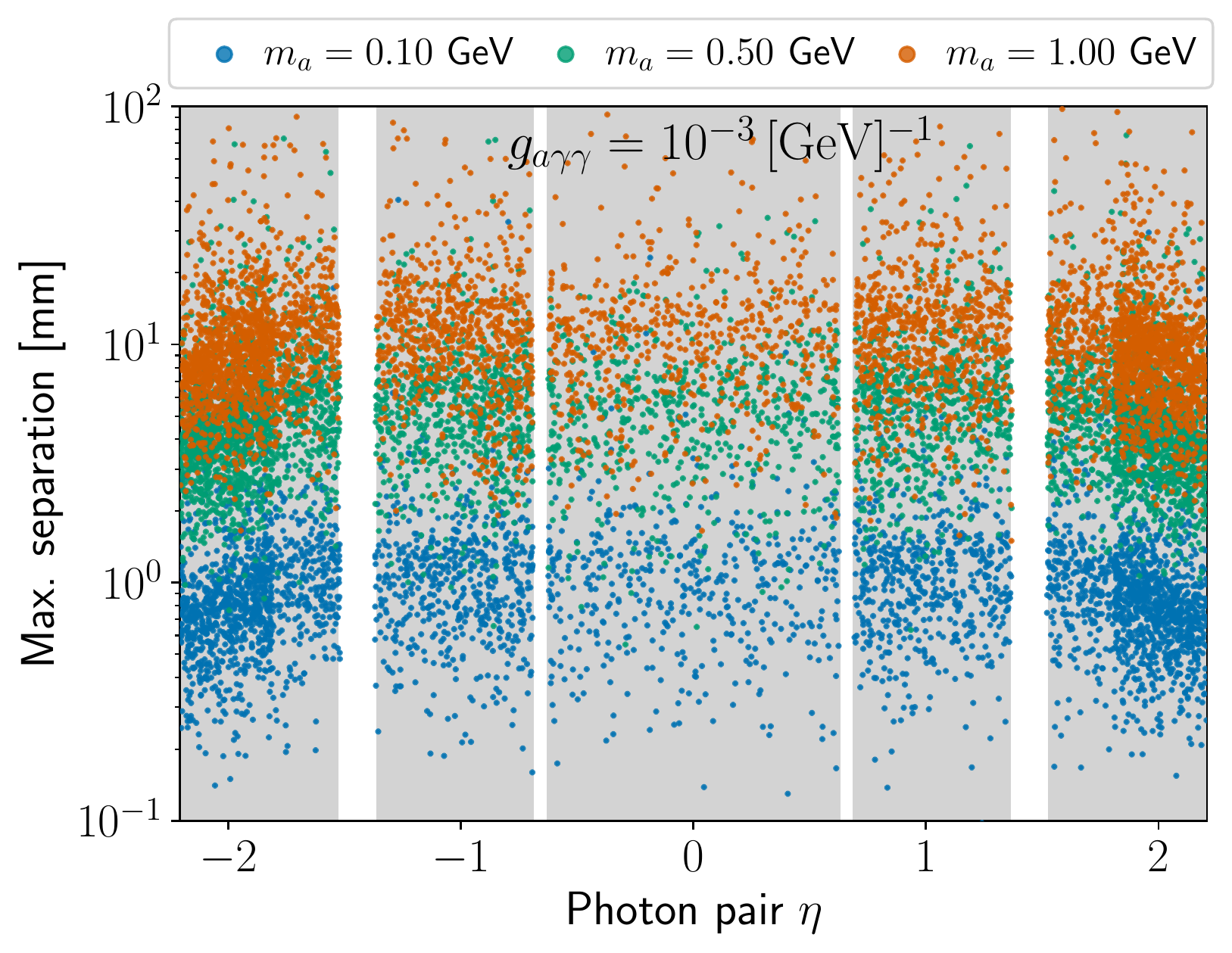}
&
    \includegraphics[width=0.49\textwidth]{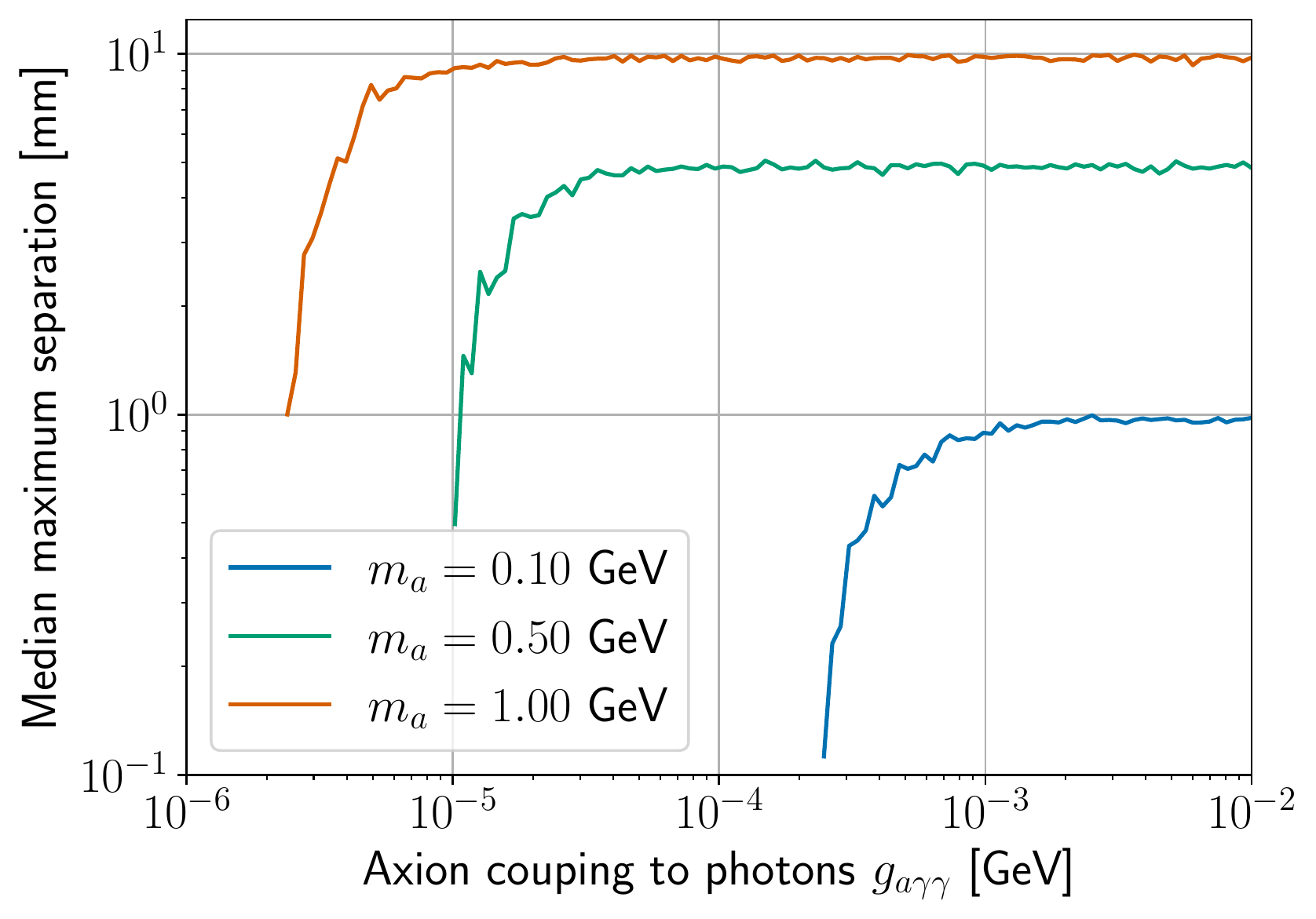}
\\
\end{tabular}
\vspace{-0.2cm}
\caption{Maximum separation between the trajectories of each photon in a photon pair originating from an ALP decay, as a function of the pseudorapidity of the photon pair (left) and the ALP coupling to photons (right).
On the left panel, the photon coupling is fixed to  $g_{a\gamma\gamma}=1\,\mathrm{TeV}^{-1}$ and each dot corresponds to one of $10^5$ simulated events for the values of the ALP mass indicated in the legend.
The grey shading indicates the coverage of the TRT, the white gaps corresponding to the barrel-endcap transition at $\eta\sim0.65$ and a region where photon conversion data is not available at $1.37<\eta<1.52$.
On the right panel, we take the median over the ALP momenta to more clearly show the trend.
}
\label{fig:separations}
\end{figure}

In order to assess the ability of the TRT to separate highly collimated converted photons, we calculate the maximum separation between the trajectories of the two photons.
This maximum separation is achieved at the edge of the TRT, just before the photons leave the inner detector.
To illustrate our results, the left panel of Fig.~\ref{fig:separations} shows the distribution of separations in the tracker for an exemplary value of the ALP-photon coupling and a few values of the ALP mass. 
Each dot represents an individual event in a series of $10^5$ simulations per value of $m_a$.
The vertical axis shows the maximum separation in the TRT of the two photons originating from a single ALP decay, while the horizontal axis displays the average pseudorapidity of those photons. Note that, for simplicity, we do not account for the bending of trajectories in the magnetic field of the detector. We do not expect that including this known effect in a more realistic simulation significantly changes the results.

The clear increase of event rate towards the edges of the TRT is mainly due to the fact that the conversion probability of the photons is larger at larger $|\eta|$, as will be discussed in Sec.~\ref{sec:conversion} and can be seen in Fig.~\ref{fig:converted_photons}.
The typical maximum separation between the photons is largely independent of $\eta$ across the TRT.
It only shows a slight decline at the edge of the endcap, which is caused by a combination of geometric factors and the fact that forwardly produced ALPS typically have a slightly larger momentum than central ones.

The separation depends most strongly on the ALP mass and its momentum, given that the angular separation between the photons is proportional to the ratio $m_a/p_a$.
For fixed values of $m_a$ and $p_a$, the separation is essentially independent of $g_{a\gamma\gamma}$ while the ALP decay length is small.
However, as $g_{a\gamma\gamma}$ decreases enough for the decay length to become larger than a few centimeters, the separation sharply decreases with decreasing $g_{a\gamma\gamma}$, as can be seen in the right panel of Fig.~\ref{fig:separations}.
This is due to the fact that in this regime the ALP decay is macroscopically displaced, and thus the photons travel a shorter distance within the TRT.
As a consequence, even if their angular separation stays constant, the physical separation between the photon trajectories at the edge of the tracker is smaller than for a prompt ALP decay.
For even longer decay lengths, the ALP decays outside of the inner detector and therefore the event does not leave any track in the TRT.
The impact of the ALP lifetime on our search strategy is further discussed in Sec.\ref{sec:displaced}.

\bigskip
To sum up the discussion so far, collimated photon pairs that convert in the tracker can be distinguished from single photon events if their separation is large enough to be resolved in the tracker.
The separation mainly depends on the mass and coupling via the amount of displacement of the vertex ($\sim p_{a}/(g^{2}_{a\gamma\gamma}m^{4}_{a})$) and the corresponding opening angle ($\sim m_{a}/p_{a}$).
While the pseudorapidity of the ALP and the geometry of the detector play a role, their effects are modest. 
For our analysis we rely on the standard ATLAS $p_T$ trigger of $140$~GeV on isolated photons which, as we will see, helps reduce the background from long-lived kaons but at the same time significantly reduces the number signal events.
This trigger requirement can potentially be relaxed if it is possible to identify a two photon conversion event or even a displaced ALP decay vertex.

\subsection{Double photon conversion in the ATLAS inner detector}\label{sec:conversion}

\begin{figure}[t]
\centering
\renewcommand{\arraystretch}{1}
\begin{tabular}[t]{cc}
	\includegraphics[width=0.472\textwidth]{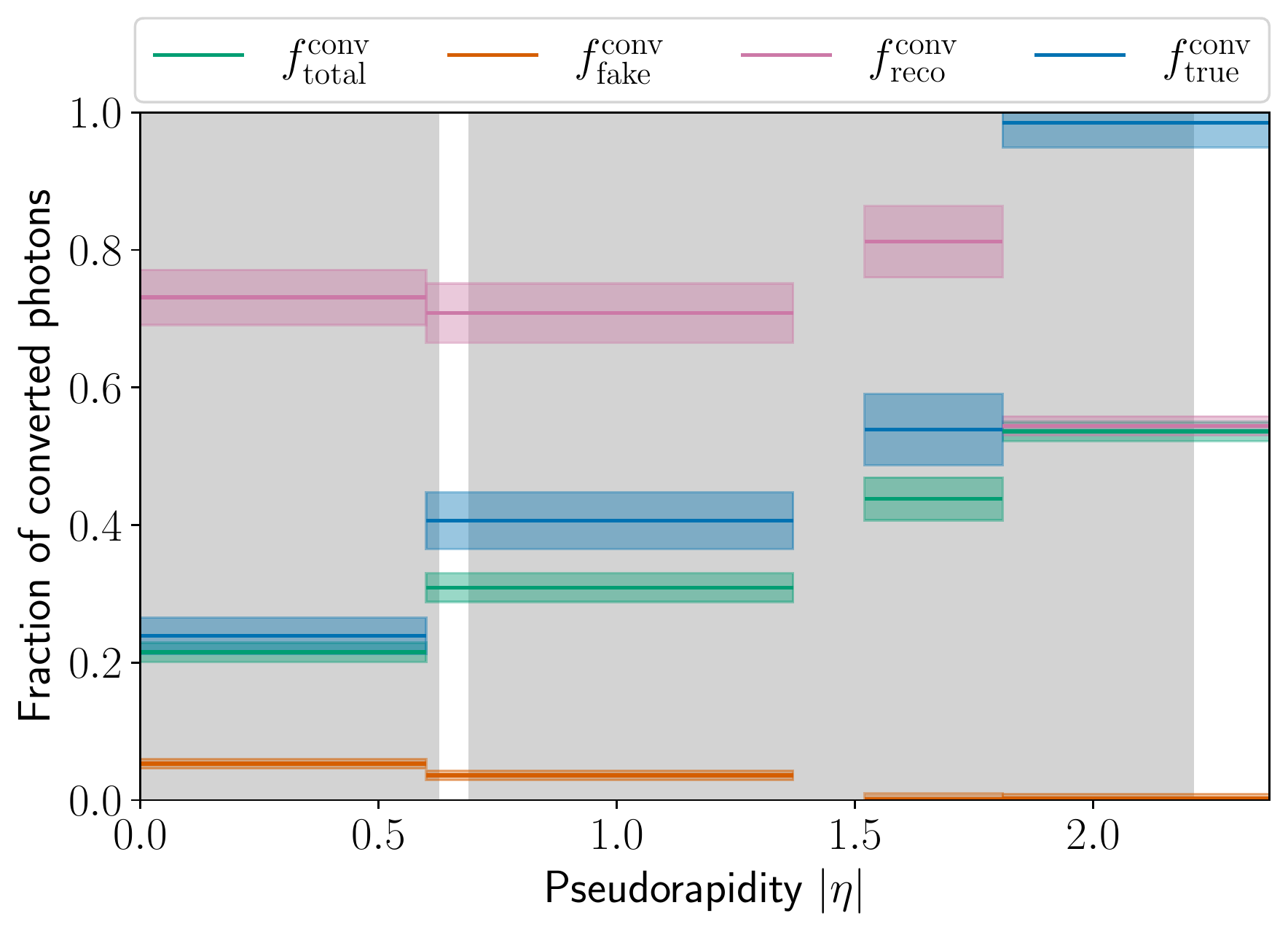}
&
    \includegraphics[width=0.472\textwidth]{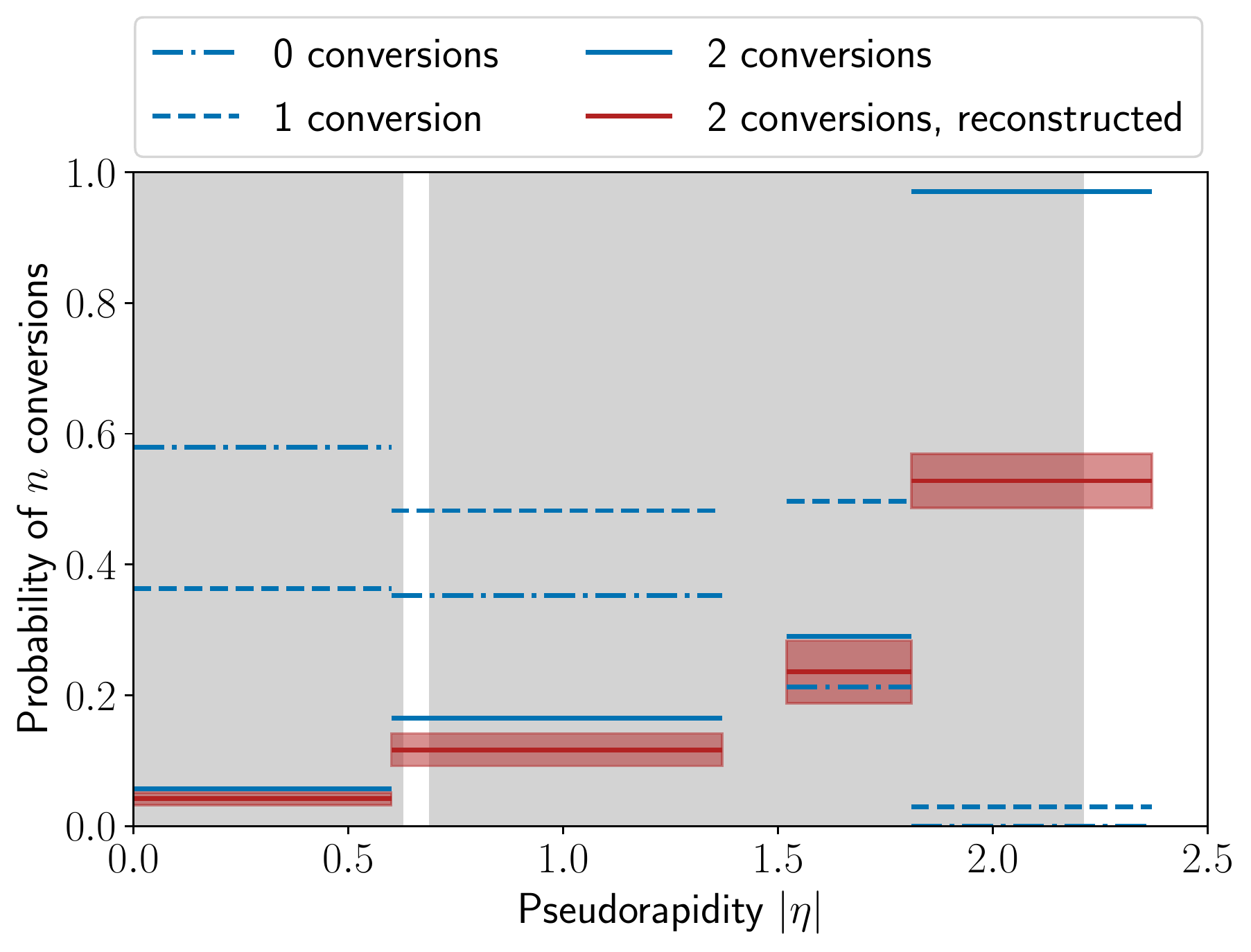}
\\
\end{tabular}
\vspace{-0.2cm}
\caption{Left panel: fraction of converted photons in the ATLAS detector in bins of $|\eta|$, as published in~\cite{ATLAS:2018fzd}.
The $f^{\mathrm{conv}}_{\mathrm{total}}$ fraction corresponds to the fraction of all photons that are reconstructed as converted, $f^{\mathrm{conv}}_{\mathrm{fake}}$ is the fraction of true unconverted photons reconstructed as converted, and $f^{\mathrm{conv}}_{\mathrm{reco}}$ the one of true converted photons reconstructed as converted. From those values, it is possible to derive $f^{\mathrm{conv}}_{\mathrm{true}}$, corresponding to the fraction of true converted photons among all photons.
Right panel: probability of $n=0,\,1,\,\mathrm{or}\,2$ conversions in a highly collimated photon pair in the ATLAS inner detector in bins of $|\eta|$. 
The blue lines show the true probability of $n$ true conversions, while the red line further takes into account the estimated reconstruction efficiency.
In both panels, the width of the shaded bands represents the uncertainty in the probabilities.
}
\label{fig:converted_photons}
\end{figure}

A photon travelling through the inner detector only leaves a track after it converts to an electron-positron pair.
The probability of conversion of a photon before it reaches the ECAL as a function of pseudorapidity in the ATLAS detector is shown in the left panel of Fig.~\ref{fig:converted_photons} using the data published in~\cite{ATLAS:2018fzd}.
The bands correspond to 1-$\sigma$ uncertainties.
This information is given in terms of three quantities: 
(i) $f^{\mathrm{conv}}_{\mathrm{total}}$, the fraction of all photons that are reconstructed as converted, 
(ii) $f^{\mathrm{conv}}_{\mathrm{fake}}$, the fraction of true unconverted photons reconstructed as converted, and
(iii) $f^{\mathrm{conv}}_{\mathrm{reco}}$, the fraction of true converted photons reconstructed as converted. 
From these quantities, we can calculate the true fraction all photons that convert,
\begin{equation}
    f^{\rm conv}_{\rm true} = \frac{f^{\rm conv}_{\rm total} - f^{\rm conv}_{\rm fake}}{f^{\rm conv}_{\rm reco} - f^{\rm conv}_{\rm fake}}.
\end{equation}
This quantity is also shown in the left panel of Fig.~\ref{fig:converted_photons}, and varies from $f^{\rm conv}_{\rm true}\simeq 0.24$ at low $\eta$ to $f^{\rm conv}_{\rm true}\simeq 0.99$ at $\eta\gtrsim 1.831$.
Note that no data is provided in the range $1.37<\eta<1.52$.
The reason why the conversion probability is larger at larger $\eta$ is simply that the photons traverse more material before reaching the ECAL.

With this, we can compute the probability for $0$, $1$, or $2$ conversions in a highly collimated photon pair as a function of pseudorapidity.
As can be seen in the right panel of Fig.~\ref{fig:converted_photons}, the fraction of events where 2 conversions occurs ranges from about $5\%$ at low $\eta$ to almost $100\%$ at the edge of the TRT.
A quantity that is more challenging to estimate is the probability for the doubly-converted photon pairs to be identified as such in the reconstruction procedure.
In the standard single-photon case, this probability is around $f^{\rm conv}_{\rm reco}\sim 0.5-0.8$ depending on $\eta$ (see left panel of Fig.~\ref{fig:converted_photons}).
Although it is likely that the reconstruction probability is lower in the double-conversion case, it is difficult to make a quantitative guess without a dedicated experimental study that is beyond our capabilities.
Thus, for our exploratory study we will take the reconstruction probability for double conversions to equal the one for single conversions, that is, $f^{\rm conv}_{\rm reco}$.
Implementing this leads to the red lines and associated uncertainties shown in the right panel of Fig.~\ref{fig:converted_photons}.

The conversion probabilities shown in Fig.~\ref{fig:converted_photons} are applicable to prompt photons, but in our model the photons originate from the ALP decay which may be displaced from the primary interaction vertex.
In order to evaluate how the conversion probability scales with the distance travelled within the inner detector, one in principle needs a detailed characterization of the material that is traversed by the particle, as is done in~\cite{ATLAS:2018fzd}.
However, for simplicity, in our estimates we take the inner detector to be homogeneous.
In this approximation, the conversion probability is constant along the photon's path through the detector, and so the total probability for it to convert if the trajectory starts at a distance $x$ from the primary interaction vertex is $f_{\rm true}^{\rm conv} (1 - x/L_{\rm TRT})$.
Here $L_{\rm TRT}=L_{\rm TRT}(\eta)$ is the total extent of the TRT in the direction of travel.
Combining this with the probability for the ALP to decay at a given distance from its production point, the conversion probability for photons originating from a long-lived ALP is
\begin{align}\label{eq:conversion_probability}
    f^{\rm conv}_{\rm true}(\ell_{\rm decay}) &= \int_{0}^{L_{\rm TRT}} \frac{\mathop{\diff x}}{\ell_{\rm decay}} \me^{-\frac{x}{\ell_{\rm decay}}}\,f^{\rm conv}_{\rm true} \left( 1 - \frac{x}{L_{\rm TRT}} \right)\nonumber \\
    &= f^{\rm conv}_{\rm true} \left[ 1 - \left( 1-\me^{-\frac{L_{\rm TRT}} {\ell_{\rm decay}}} \right) \frac{\ell_{\rm decay}}{L_{\rm TRT}} \right],
\end{align}
with $\ell_{\rm decay}$ the ALP decay length given in Eq.~\eqref{eq:l_decay}.
As expected, the conversion probability is strongly suppressed for $\ell_{\rm decay} \gtrsim L_{\rm TRT}$.
In our simulations, we randomly decide whether the photons convert and generate the location for the conversion vertices based on the above probability.

\subsection{Displaced ALP decay vertex}\label{sec:displaced}
ALPs are not the only particle that can produce a highly collimated photon pair at the LHC.
The decay of other states, in particular of light hadronic resonances, can mimic the final state that our proposed search strategy targets.
Most importantly, the LHC produces a copious amount of boosted neutral particles that decay preferentially into a diphoton final state.
This results in an enormous background that needs to be suppressed in order to render our search strategy viable.
Fortunately, there is one property of the ALPs that can help with that: their long lifetime.

As highlighted by Eq.~\ref{eq:l_decay}, the decay length of a sub-GeV ALP with couplings below the TeV$^{-1}$ range is macroscopic on collider scales when taking into account the boost factor of the ALP.
Neutral pions and other light hadronic resonances that decay into $\gamma\gamma$ are much shorter-lived (we briefly discuss those in section~\ref{sec:reducible}). The only exception to this are $K_L$, which will be discussed in the section~\ref{sec:irreducible_backgrounds}. Thus, the identification of a non-zero displacement for the $a\rightarrow\gamma\gamma$ decay vertex with respect to the primary interaction vertex (IV) would allow to immediately reject all those sources of background.

\begin{figure}[t]
\centering
\setlength{\tabcolsep}{10pt}
\renewcommand{\arraystretch}{1}
\begin{tabular}{cc}
		\includegraphics[width=0.455\textwidth]{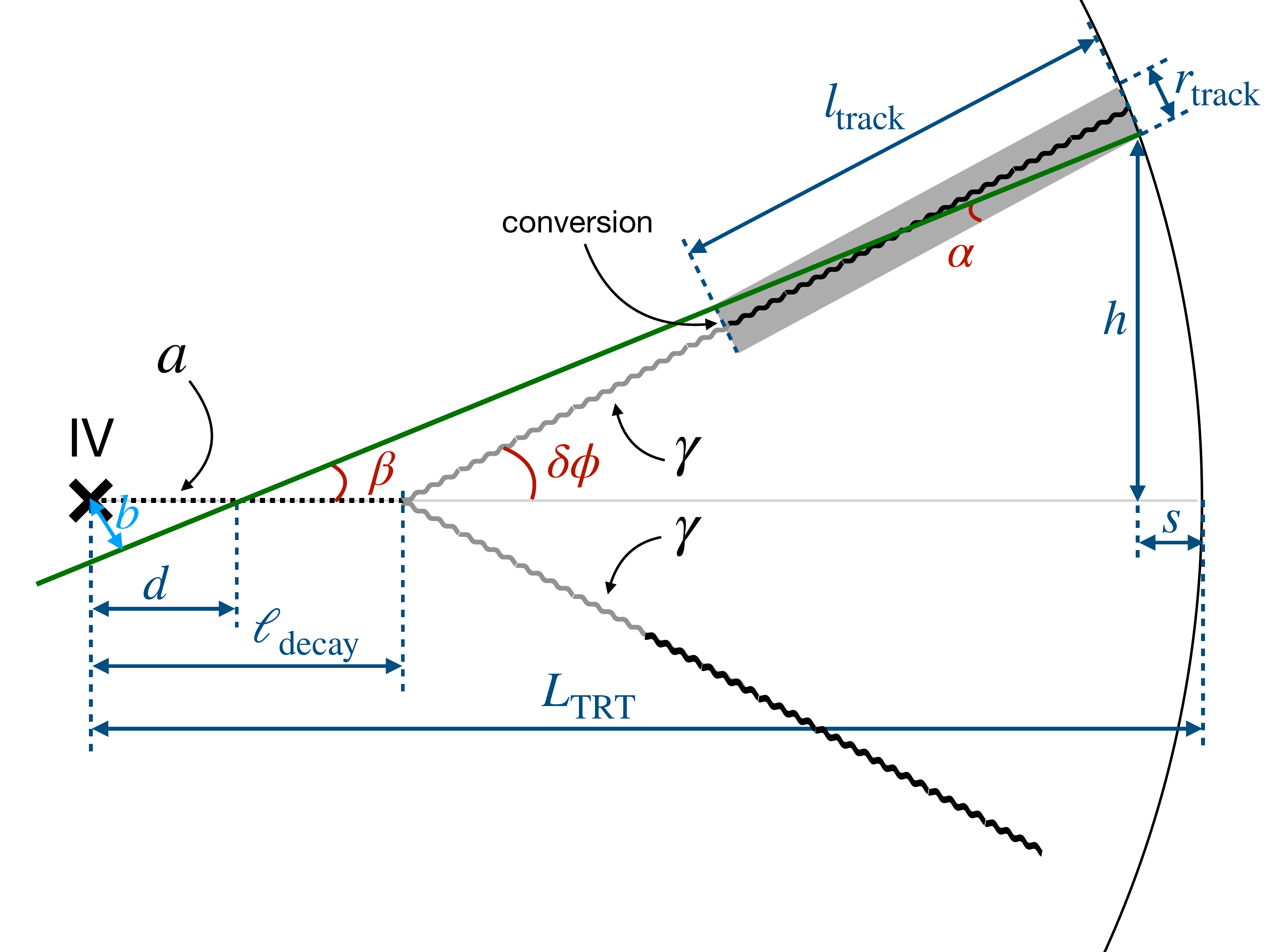}
&
		\includegraphics[width=0.455\textwidth]{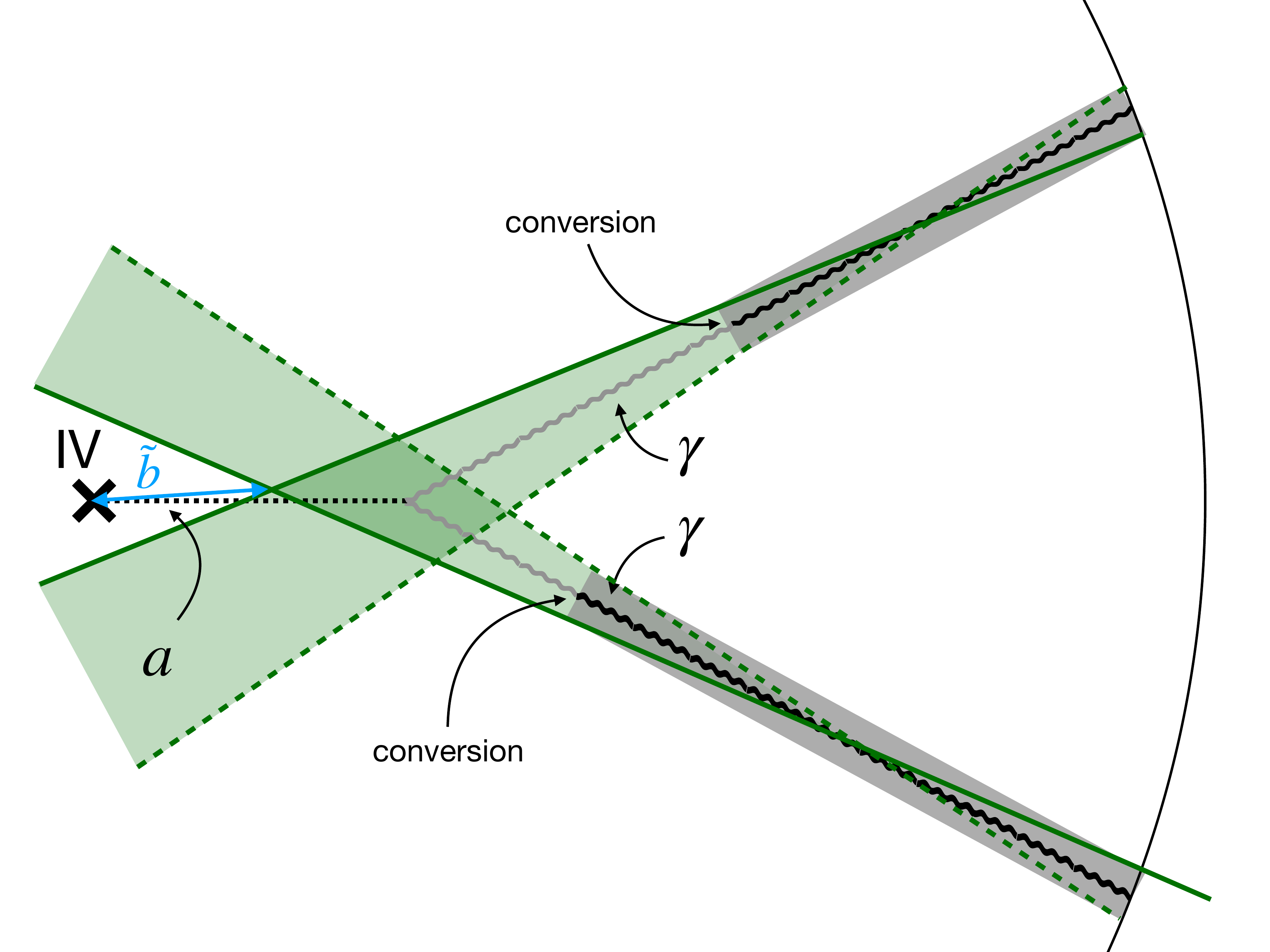}
\\
\end{tabular}
\vspace{-0.2cm}
\caption{Sketches showcasing the geometry related to the displaced decay of the axion into two photons ($a \rightarrow \gamma\gamma$) at the ATLAS inner detector.
Everything is projected on the transverse plane.
As can be seen in the left panel, the observable transverse impact parameter $b$ of a photon's trajectory with respect to the interaction vertex (IV) critically depends on the collimation angle between the decay photons, the length of the photon track $l_{\rm track}$ after it converts in the TRT, and the resolution $r_{\rm track}$ with which the photon trajectory can be determined.
The right panel showcases that the determination of the displaced vertex is improved if both photons can be distinguished as displaced.
Note that these sketches do not represent the typical conditions of the events of interest here, where $\delta\phi$ is much smaller. See the text for more details.}
\label{fig:displaced_vertex_geometry}
\end{figure}

In practice, the ability to measure the displacement of the axion decay is greatly hindered by the fact that its decay products are highly collimated.
This can be understood by looking at Fig.~\ref{fig:displaced_vertex_geometry}, which sketches the geometry of the process at hand.
Given the inherent uncertainties in determining the location of the IV along the beam axis, we only consider displacements in the transverse plane.

Let us first consider the situation that is sketched in the left hand panel of Fig.~\ref{fig:displaced_vertex_geometry} and use only information from a single photon to determine whether the vertex is displaced. In this case the smallest possible impact parameter consistent with the measured trajectory determines the (minimal) displacement.
The axion is produced at the IV, and travels a distance $\ell_{\rm decay}$ before decaying into $\gamma\gamma$.
We denote by $\delta\phi$ the difference in azimuthal angle between then trajectory of the ALP and that of the photon.
As can be gathered from Fig.~\ref{fig:DeltaR_distribution}, $\delta\phi$ is typically a small angle, below $0.01$~rad for sub-GeV ALP masses.
The photon only leaves a visible track in the inner detector once it converts to an $e^\pm$ pair: we denote the length of such track by $l_{\rm track}$.
The accuracy with which ATLAS can establish the trajectory of the photon is determined by the position resolution of the TRT, which is around $0.1$~mm~\cite{ATLAS:2017jwu,2017NIMPA.845..257M}.
In our sensitivity estimations, we parametrize the precision to determine the trajectory in terms of the quantity $r_{\rm track}$ defined in Fig.~\ref{fig:DeltaR_distribution}.
As a benchmark, we take it to be comparable to the spatial resolution of the TRT, $r_{\rm track}\sim 0.1$~mm.
A better resolution may, however, be achievable for tracks that are sufficiently long.
As we will see, a smaller value of $r_{\rm track}$ results in a much better displaced vertex discrimination power.

In order to estimate the ability to distinguish a displacement between the IV and the axion decay vertex, we identify the most extreme possible reconstructed photon trajectory as the dark green line in Fig.~\ref{fig:displaced_vertex_geometry}.
This line differs from the actual photon trajectory by an angle $\alpha$ given by
\begin{equation}\label{eq:alpha}
    \tan\alpha = \frac{r_{\rm track}}{l_{\rm track}},
\end{equation}
and thus makes an angle $\beta = \delta\phi - \alpha$ with the ALP trajectory.
The distance $d$ between the IV and the intersection between this most extreme trajectory and the ALP trajectory is given by
\begin{equation}\label{eq:d}
    d = L_{\rm TRT} - s - \frac{h}{\tan\beta}.
\end{equation}
Here, we have defined the lengths $h$ and $s$ as indicated in Fig.~\ref{fig:displaced_vertex_geometry}.
Elementary trigonometry allows to write them as functions of known quantities.
For $h$, we find
\begin{equation}
    h = \tan\mathop{\delta\phi} \left( L_{\rm TRT} - \ell_{\rm decay} - \frac{r_{\rm track}}{2\sin\mathop{\delta\phi}} - s \right),
\end{equation}
while $s$ corresponds to the smallest of the two roots of the quadratic equation
\begin{equation}
    \left[ 1 + \tan^2\mathop{\delta\phi} \right] s^2 - 2\left[ L_{\rm TRT} + 4\left( L_{\rm TRT} - \ell_{\rm decay} - \frac{r_{\rm track}}{2\sin\mathop{\delta\phi}} \right)\tan^2\mathop{\delta\phi} \right] s + 4\tan^2\mathop{\delta\phi} = 0.
\end{equation}
Once $d$ is determined, the impact parameter of the most extreme photon reconstructed trajectory with respect to the IV (i.e. the distance of closest approach, see Fig.~\ref{fig:displaced_vertex_geometry}) can be calculated as
\begin{equation}
    b = d \sin\beta.
\end{equation}
For small values of $\delta\phi$ and $\alpha$, we can take $s\approx 0$ to obtain the approximate expression 
\begin{equation}\label{eq:b_approx}
    b \approx \mathop{\delta\phi} \ell_{\rm decay} - \alpha L_{\rm TRT}.
\end{equation}
Although for our numerical simulations we use the exact expression for $b$, Eq.~\eqref{eq:b_approx} allows to qualitatively understand our results.
First, $b<0$ signals that the IV is within the possible reconstructed photon trajectories, meaning that the ALP decay cannot be tagged as displaced.
As $\ell_{\rm decay} \leq L_{\rm TRT}$ for the ALP to decay within the tracker, $b>0$ necessarily requires that $\alpha \leq \mathop{\delta\phi}$.
In fact, the shorter the ALP decay length is, the larger $\delta\phi$ needs to be in order to compensate it.
Note that Eq.~\eqref{eq:b_approx} is not applicable when $L_{\rm TRT} - \ell_{\rm decay} \sim r_{\rm track}$, as in that case $l_{\rm track}\sim r_{\rm track}$ and Eq.~\eqref{eq:alpha} implies that $\alpha$ is not a small angle anymore.

Equation~\eqref{eq:b_approx} also gives an idea of the order of magnitude of the impact parameter.
Given that decays within the TRT require $\ell_{\rm decay}\lesssim 1$~m and that sub-GeV ALPS have $\mathop{\delta\phi}\lesssim 10^{-2}$ (see Fig.~\ref{fig:DeltaR_distribution}), we conclude that in typical ALP decays we will deal with impact parameters as small as $b\sim 10$~cm.
Although that seems small, long-lived particle searches at ATLAS have achieved sub-cm accuracy in distinguishing displaced decays~\cite{ATLAS:2017tny}.
It is thus reasonable to assume that impact parameters as small as $1$~cm could be distinguished in our case.
That said and to be conservative, in our benchmark scenario we require a displacement of at least $b>10$~cm, which we only relax down to $b>1$~cm for our most optimistic projections. 

As can be seen in the right panel of Fig.~\ref{fig:displaced_vertex_geometry}, the ability to discern a displaced decay improves considerably if the reconstructed trajectories of both photons from the ALP decay are taken into account.
In that case, the possible location of the displaced vertex can be narrowed down to the intersection region of the two cones shaded in green in Fig.~\ref{fig:displaced_vertex_geometry}.
The point in that region closest to the IV defines the minimal reconstructed displacement $\tilde{b}$, which is given by
\begin{equation}\label{eq:b_2photons}
    \tilde{b} = \sqrt{\sin^2\beta_1 x^2 + (d_1+x)^2},
\end{equation}
where
\begin{equation}
    x = \frac{\sin\beta_2}{\sin\beta_1+\sin\beta_2}(d_2-d_1).
\end{equation}
Here, $d_1$ and $d_2$ correspond to the individual distances given by Eq.~\eqref{eq:d} for each photon with the convention that $d_2>d_1$ ($\beta_1$ and $\beta_2$ are the corresponding angles as defined in Fig.~\ref{fig:displaced_vertex_geometry}).
The expression Eq.~\eqref{eq:b_2photons} is only valid if $d_1$ and $d_2$ are both positive.
For very collimated photons, Eq.~\eqref{eq:b_2photons} simplifies to
\begin{equation}\label{eq:b_2photons_approx}
    \tilde{b} \approx \frac{1}{\beta_1 + \beta_2} \left( \mathop{\delta\phi_1}\ell_{\rm decay,\,1} + \mathop{\delta\phi_2}\ell_{\rm decay,\,2} \right) - L_{\rm TRT} \left( \frac{\mathop{\delta\phi_1}+\mathop{\delta\phi_2}}{\beta_1+\beta_2} - 1 \right),
\end{equation}
where $\beta_i = \mathop{\delta\phi_i}-\alpha_i$.
Although $\mathop{\delta\phi_1}\simeq \mathop{\delta\phi_2} \simeq \mathop{\delta\phi_{\gamma\gamma}}/2$, $\alpha_1$ and $\alpha_2$ can be very different depending on where each of the photons converts.
Compared to Eq.~\eqref{eq:b_approx}, Eq.~\eqref{eq:b_2photons_approx} is not suppressed by a factor of the small angles.
Thus, when both photon trajectories can be reconstructed as displaced, the reconstructed displacement for the ALP decay vertex can be significantly larger than a few cm even for light ALPs.

In our simulations, we only consider as signal candidates those events where the ALP decay vertex can be identified as displaced based on the reconstruction using both converted photons.
At the price of a significant reduction in the signal event production cross-section, this allows to reject all known backgrounds except for one: long-lived Kaons, which have a macroscopic decay length and a small but nonzero branching ratio into photon pairs~\cite{ParticleDataGroup:2020ssz}.
We make an estimation of the impact of that irreducible background in the next section.

\subsection{Irreducible background from $K_L$ }\label{sec:irreducible_backgrounds}

The long-lived state in the neutral Kaon pair---$K_L$---has a lifetime $\tau_{K_L}\simeq 5\times 10^{-8}$~s and a branching ratio to photons $\mathrm{Br}(K_L\rightarrow\gamma\gamma)\simeq 5\times 10^{-4}$.
This means that $K_L$ can produce the same signal that we target in our search strategy.
It is therefore crucial to estimate the rate at which this background process occurs at the LHC.

\begin{figure}[t]
\centering
\renewcommand{\arraystretch}{1}
\begin{tabular}[t]{cc}
	\includegraphics[width=0.49\textwidth]{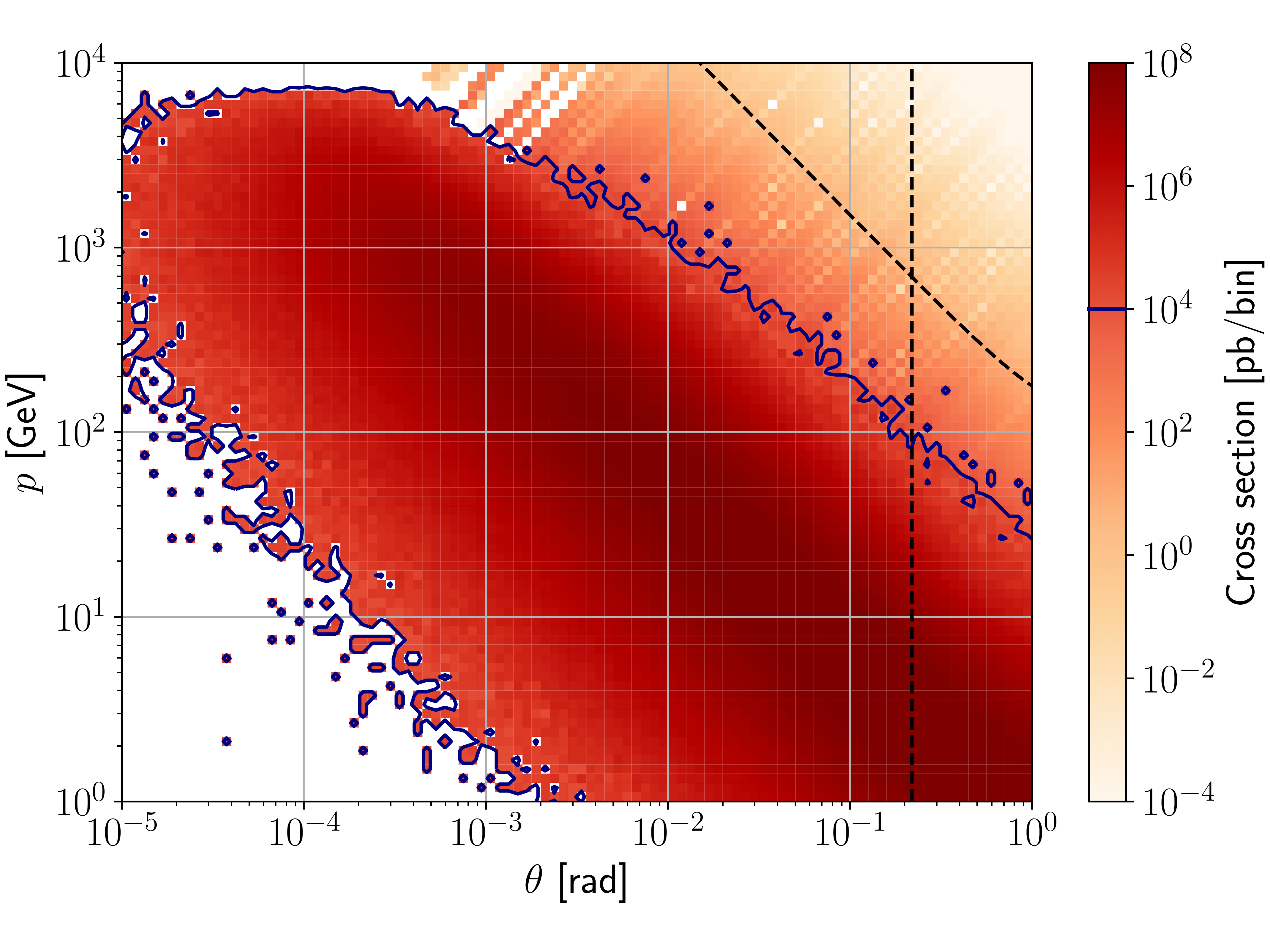}
&
    \includegraphics[width=0.47\textwidth]{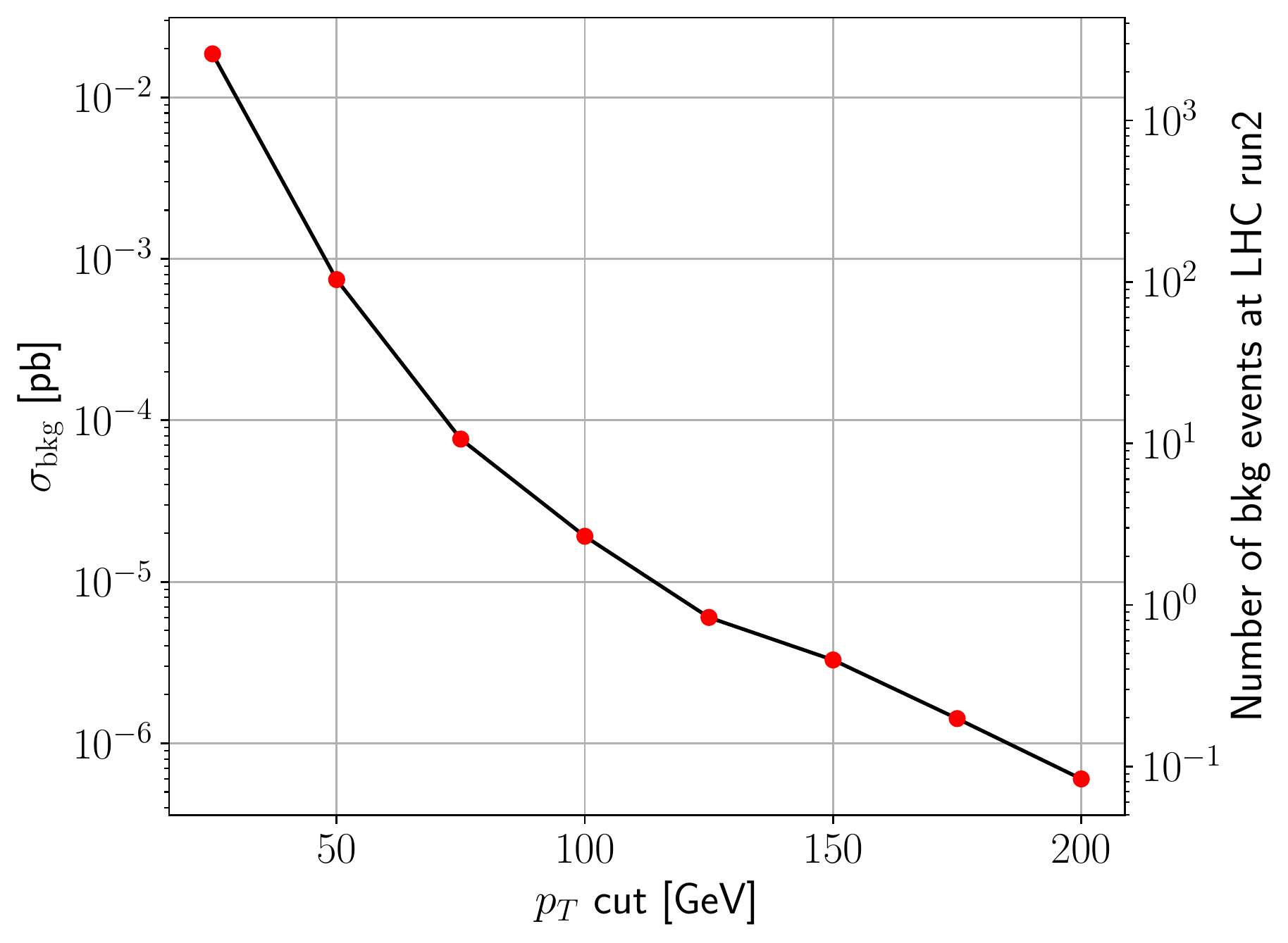}
\\
\end{tabular}
\vspace{-0.2cm}
\caption{Characterization of the irreducible background coming from $K_L\rightarrow \gamma\gamma$ events.
On the left panel and delimited by a blue line, we show the $\theta$ vs $p$ distribution of $K_L$ produced at the LHC as calculated in~\cite{Kling:2021fwx}, which has a resolution of $10^4$~pb/bin.
We also show our extrapolation of this distribution to the signal region for our search, above and to the right of the black dashed lines.
The right panel displays the cross section of $K_L\rightarrow\gamma\gamma$ events in the signal region as a function of the cut in the transverse momentum of the final state photons.
Note that this constitutes an upper limit to the number of background events that pass all of our signal selection criteria.
For easy reference, we also display the corresponding number of background events at the run 2 of the LHC.}
\label{fig:KL_background}
\end{figure}

The first step is to characterize the angular and momentum distributions of the long-lived neutral kaons produced in $pp$ collisions at $14$~TeV center-of-mass energy.
For that, we make use of the \texttt{FORESEE} package~\cite{Kling:2021fwx}, which provides such distributions for a number of hadronic states, including $K_L$.\footnote{Since $K_{L}$ decays to two photons, it features an effective coupling to two photons $g_{K_{L}\gamma\gamma}$. Kaons produced through vector boson fusion via this coupling have a similar distribution to that of the ALPs. However, the effective coupling is very small, $g_{K_{L}\gamma\gamma}\lesssim 10^{-8}\,{\rm GeV}^{-1}$, and this source of background can be safely neglected.}
The left panel of Fig.~\ref{fig:KL_background} shows the two-dimensional $\theta-p$ distribution of $K_L$.
Bins outside of the dark blue line have cross-sections below the $10^{4}$~pb threshold used for the simulation.
The distribution is fairly forward, especially for high momentum kaons.
The dashed vertical line shows the angular extend of the TRT, which extends down to $\eta\simeq 2.2$ or $\theta\simeq0.22$~rad.
The dashed diagonal line corresponds to a cut $p_T>150$~GeV.
As can be seen from the figure, the signal region for our search corresponds to the upper right corner of the $\theta-p$ plane, where no data for the $K_L$ production cross-section is available due to the lack of resolution of the simulations~\cite{Kling:2021fwx}.
To overcome this, we extrapolate the results of~\cite{Kling:2021fwx} to our signal region using a simple two-dimensional power law fit.
The result corresponds to the colored area outside the blue line in the left panel of Fig.~\ref{fig:KL_background}.

Once the production cross-section for $K_L$ in the $\theta-p$ signal region is obtained, we need to estimate the amount of  $K_L\rightarrow\gamma\gamma$ decays that are expected to leave tracks in the ATLAS TRT.
First, there is a suppression of the rate by the small branching fraction $\mathrm{Br}(K_L\rightarrow\gamma\gamma)\simeq 5\times 10^{-4}$.
Second, only a few $K_L$ decay within the TRT due to their large decay length, 
\begin{equation}
    \ell_{K_L} \simeq 4.6\,\mathrm{km}\,\frac{p}{150\,\mathrm{GeV}}.
\end{equation}
Given that the TRT extends to a maximum distance of $\sim 1$~m from the IV, only a $\sim 2\times 10^{-4}$ fraction of the kaons decay within the tracker.
A further suppression factor is due to the photon conversion probability as given by Eq.~\eqref{eq:conversion_probability}.
After these considerations are taken into account, we can estimate an upper limit for the cross section of $K_L$ background events in the signal region.
This is shown in the right panel of Fig.~\ref{fig:KL_background} as a function of the lower $p_T$ requirement to define the signal region.
For easy reference, we show the number of background events corresponding to each value of the cross section for the $139\,\mathrm{fb}^{-1}$ integrated luminosity of the second observation run of the LHC (the third observation run is expect to double the current integrated luminosity, while the HL-LHC is planned to further boost it by a factor of 10).
The background cross section depends strongly on the $p_T$ cut applied, and the expected number of background events at the run 2 of the LHC is below $1$ for a $p_T$ above $\sim 120$~GeV.

So far, we have not imposed any requirement on the collimation, i.e. the $\Delta R$, of the two photons arising from the $K_L$ decay, which would result in a further suppression factor. 
This is however not easy to estimate without a full simulation with the sufficient statistics to populate the signal region delineated in the left panel of Fig.~\ref{fig:KL_background}.
As this task is beyond the scope of the present work, we will take the results shown in the right panel of Fig.~\ref{fig:KL_background} as an upper limit to the $K_L\rightarrow\gamma\gamma$ background cross section.
Furthermore, the photon isolation criteria would also reject a significant fraction of $K_L$ produced within QCD jets.

We thus conclude that it is reasonable to expect our search to be background free at the LHC run 2 (and possibly LHC run 3) as long as the $p_T$ cut placed on the photon pair is at least $\sim 120$~GeV.
In particular, a search using events recorded with the current $140$~GeV cut for isolated photons would satisfy the background-free requirement.

We finish this subsection by noting that care also needs to be taken of other reducible but nevertheless potentially dangerous backgrounds.
Among others, these include decays where other particles are misidentified as photons or where one (or several) particles are lost in the reconstruction.
A characterization of these sources of background is beyond the scope of the present theory-level analysis.

\subsection{Reducible background from $\pi^{0}$}\label{sec:reducible}
\begin{figure}[t]
\centering
\renewcommand{\arraystretch}{1}
\begin{tabular}[t]{cc}
	\includegraphics[width=0.49\textwidth]{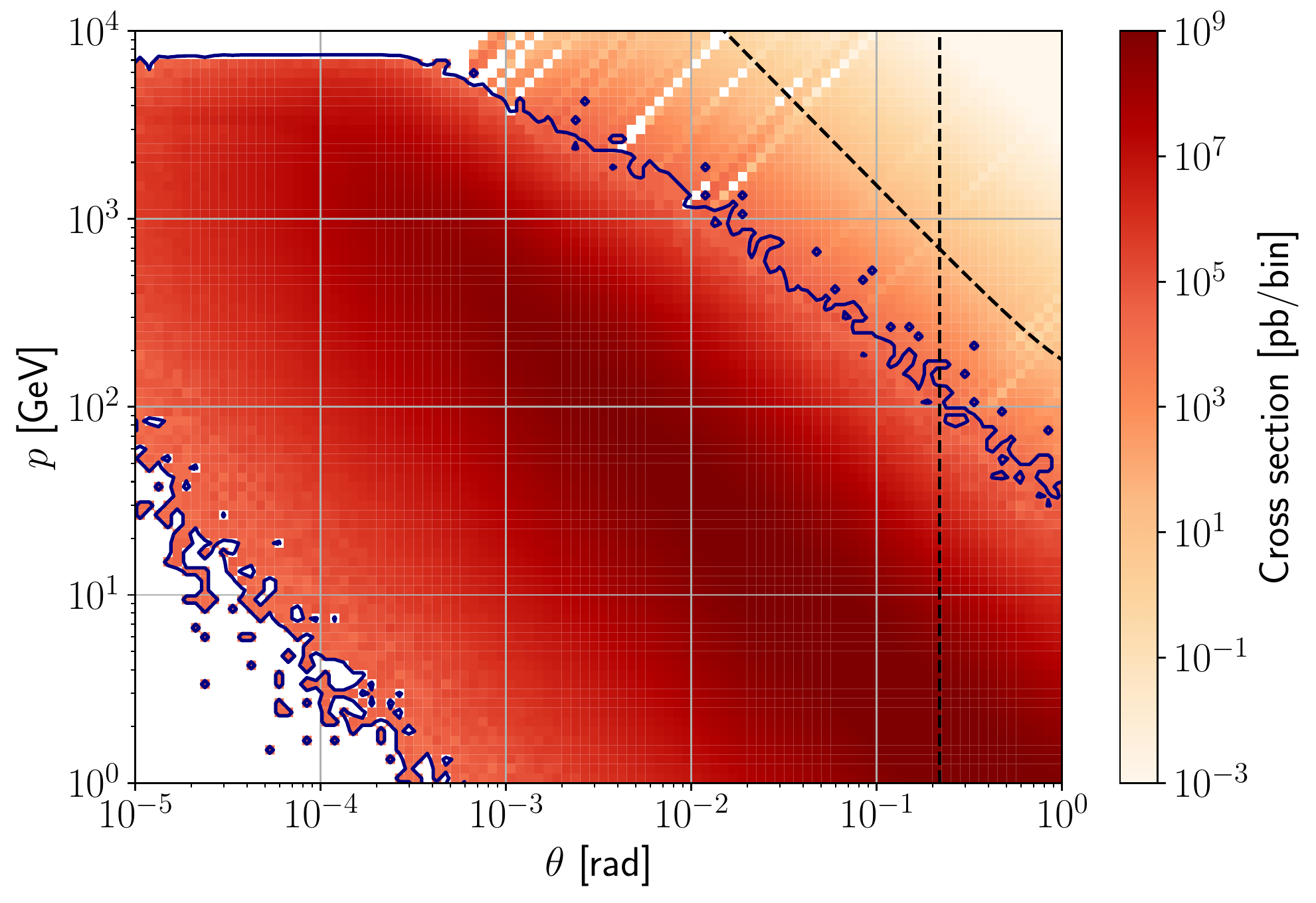}
&
    \includegraphics[width=0.47\textwidth]{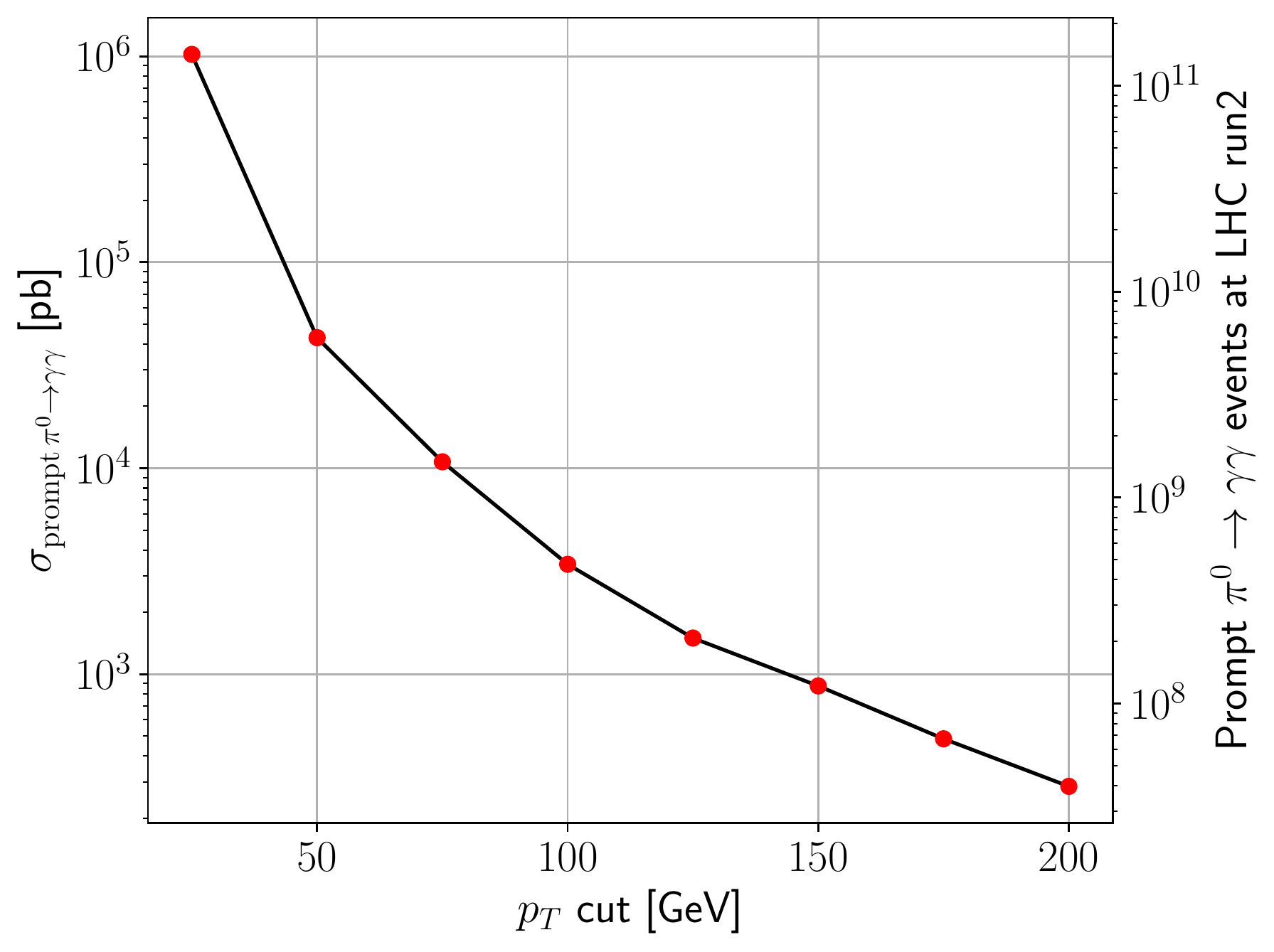}
\\
\end{tabular}
\vspace{-0.2cm}
\caption{Characterization of the prompt $\pi^0$ production.
On the left panel and delimited by a blue line, we show the $\theta$ vs $p$ distribution of $\pi^0$ produced at the LHC as calculated in~\cite{Kling:2021fwx}, which has a resolution of $10^4$~pb/bin.
We also show our extrapolation of this distribution to the momentum and pseudorapidity region of interest for our search, above and to the right of the black dashed lines.
The right panel displays the cross section of $\pi^0\rightarrow\gamma\gamma$ events in the $p$ and $\eta$ region as a function of the cut in the transverse momentum of the final state photons.
For easy reference, we also display the corresponding number of background events at the run 2 of the LHC.
Note that due to the short $\pi^0$ lifetime, the $\pi^0$ background can be rejected by requiring a minimum displacement of the signal event decay vertices.}
\label{fig:pi0_background}
\end{figure}

In addition to the $K_L$ background one may also worry about the even larger number of neutral pions that are produced at the LHC.
However, the pion lifetime is much shorter and amounts to a decay length
\begin{equation}
\ell_{\pi^{0}}\simeq 3\times 10^{-5}{\rm m}\left(\frac{p}{150\,{\rm GeV}}\right).
\end{equation}
Thus, requiring a displacement of the vertex by at least $1{\rm cm}$ should be sufficient to ensure that essentially no pions survive this cut.
Nevertheless, this statement relies on the rate of misreconstruction of the original pion decay vertex being rare enough such that this does not constitute an experimental background. 
Estimating such a rate requires a dedicated experimental study. 
At this stage, we can estimate the required rejection factor, which gives an idea of the necessary quality of the reconstruction algorithm.
To do this we once more use the spectra provided by~\cite{Kling:2021fwx} and perform a similar extrapolation to larger momenta as in the case of Kaons.
The resulting distribution is shown in Fig.~\ref{fig:pi0_background}.
We thus see that for a transverse momentum cut of $p_{T}\geq 150\,{\rm GeV}$, we require a rejection factor
\begin{equation}
    R^{\rm reject}_{\pi^{0}} \sim 10^{8}
\end{equation}
to have a background-free search with the run-2 luminosity.
For a minimum displacement of the decay vertex of $10\,{\rm cm}$ this, naively, seems achievable. 
For smaller displacements around $1\,{\rm cm}$, though, achieving such a low misreconstruction rate might be more challenging, since this minimum displacement corresponds to decays occurring inside the beampipe.
The isolation requirements also contribute to achieving the required rejection factor, since the pions produced within QCD jets should be rejected at that stage.
Once more, a detailed experimental study would be necessary to make a quantitative statement in this regard.

\section{Results}\label{sec:results}

\begin{figure}[t!]
\centering
\setlength{\tabcolsep}{0pt}
\renewcommand{\arraystretch}{1}
\begin{tabular}{ccc}
		\label{fig:vary_pT_cut}
		\includegraphics[width=0.49\textwidth]{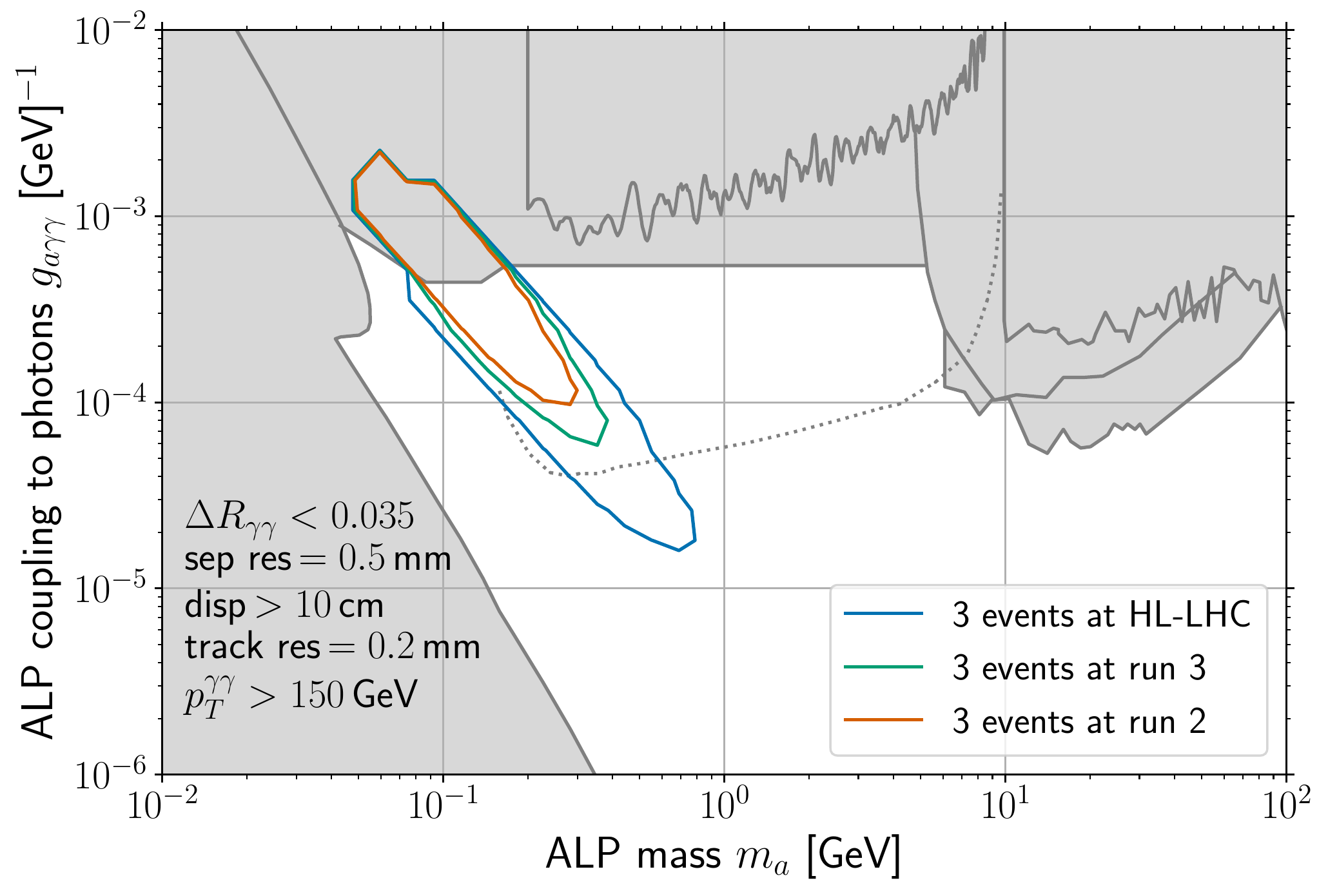}
		\label{fig:vary_N_events}
&
        \includegraphics[width=0.49\textwidth]{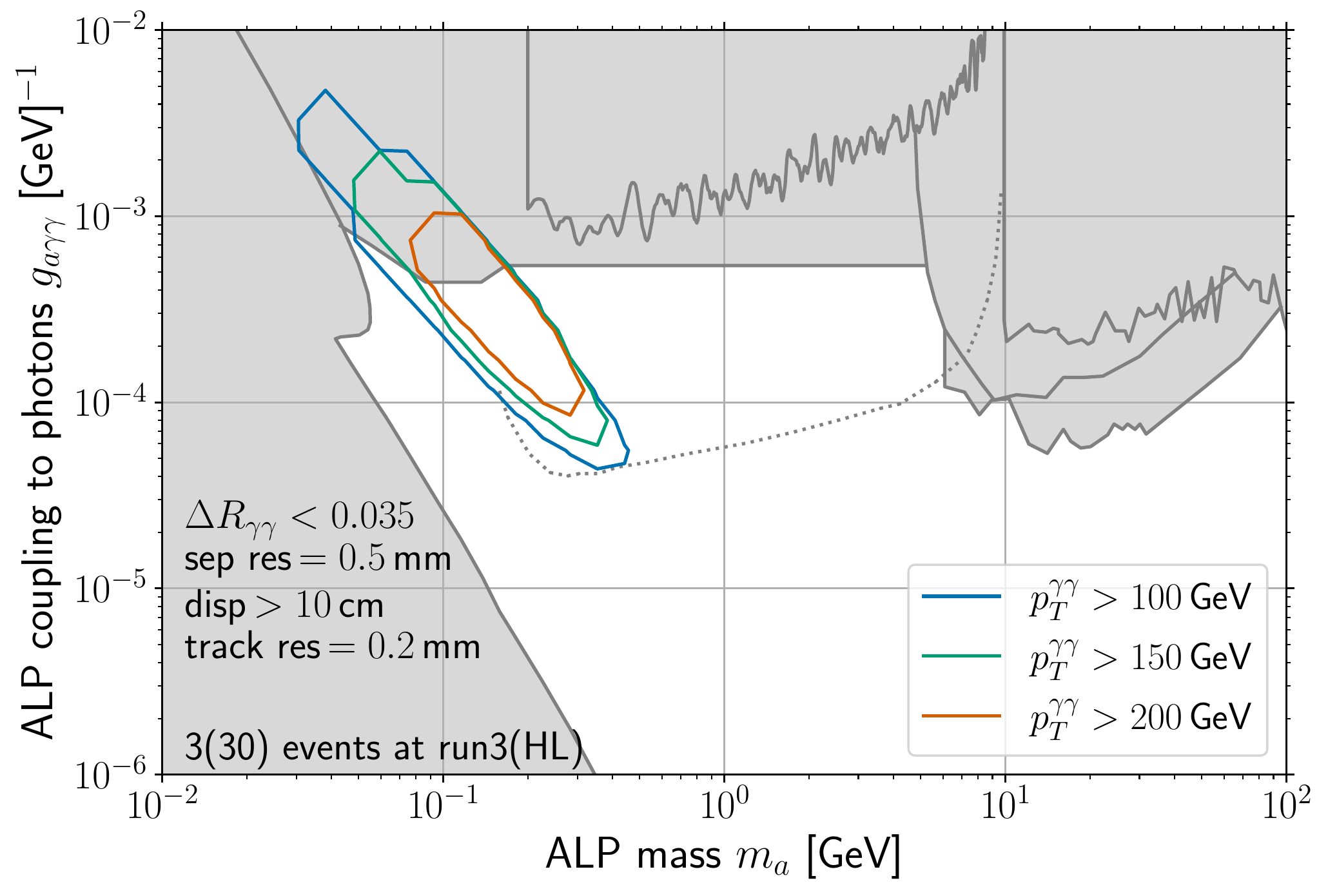}
\\
		\label{fig:vary_TRT_resolution}
		\includegraphics[width=0.49\textwidth]{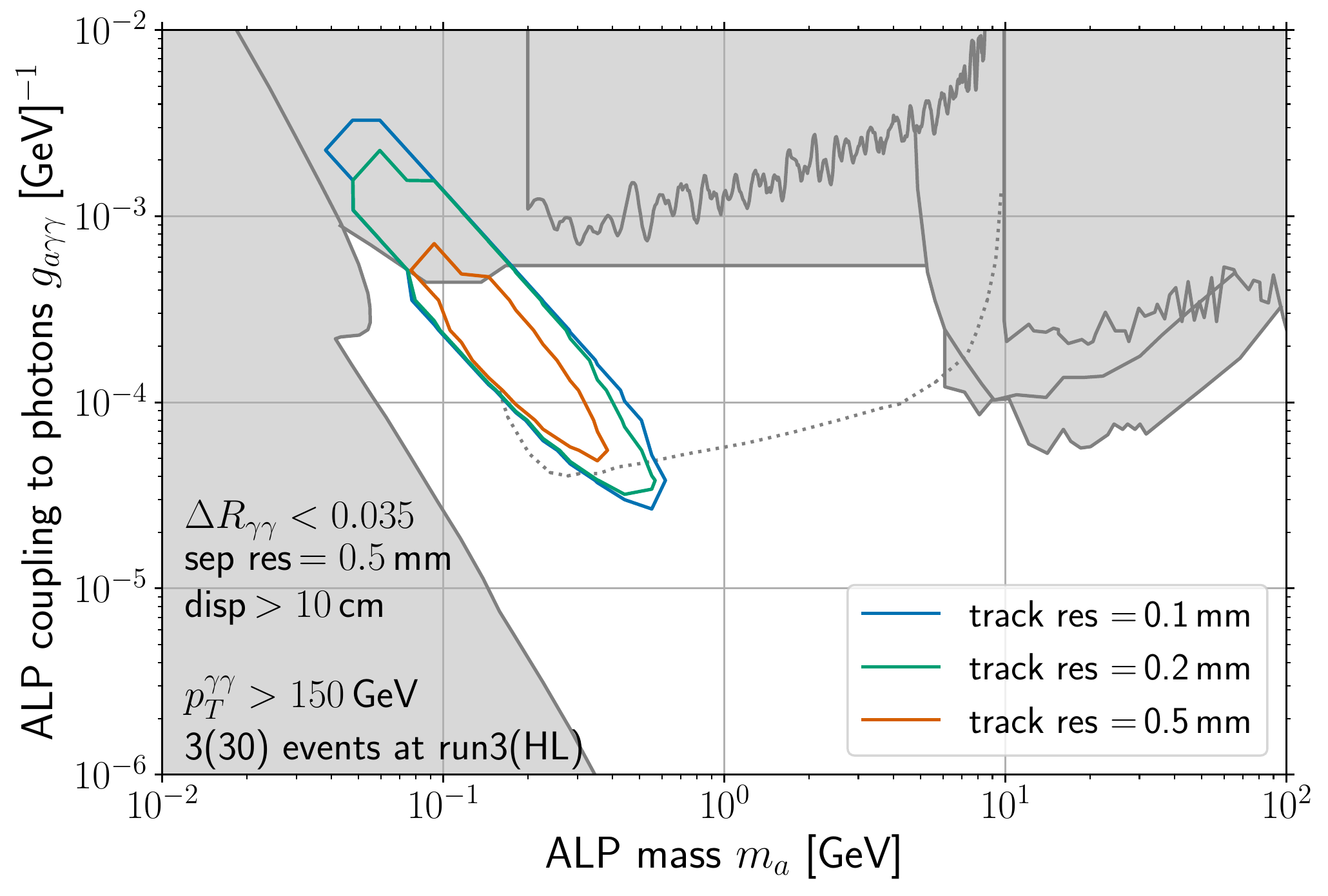}
&
		\label{fig:vary_vertex_displacement}
		\includegraphics[width=0.49\textwidth]{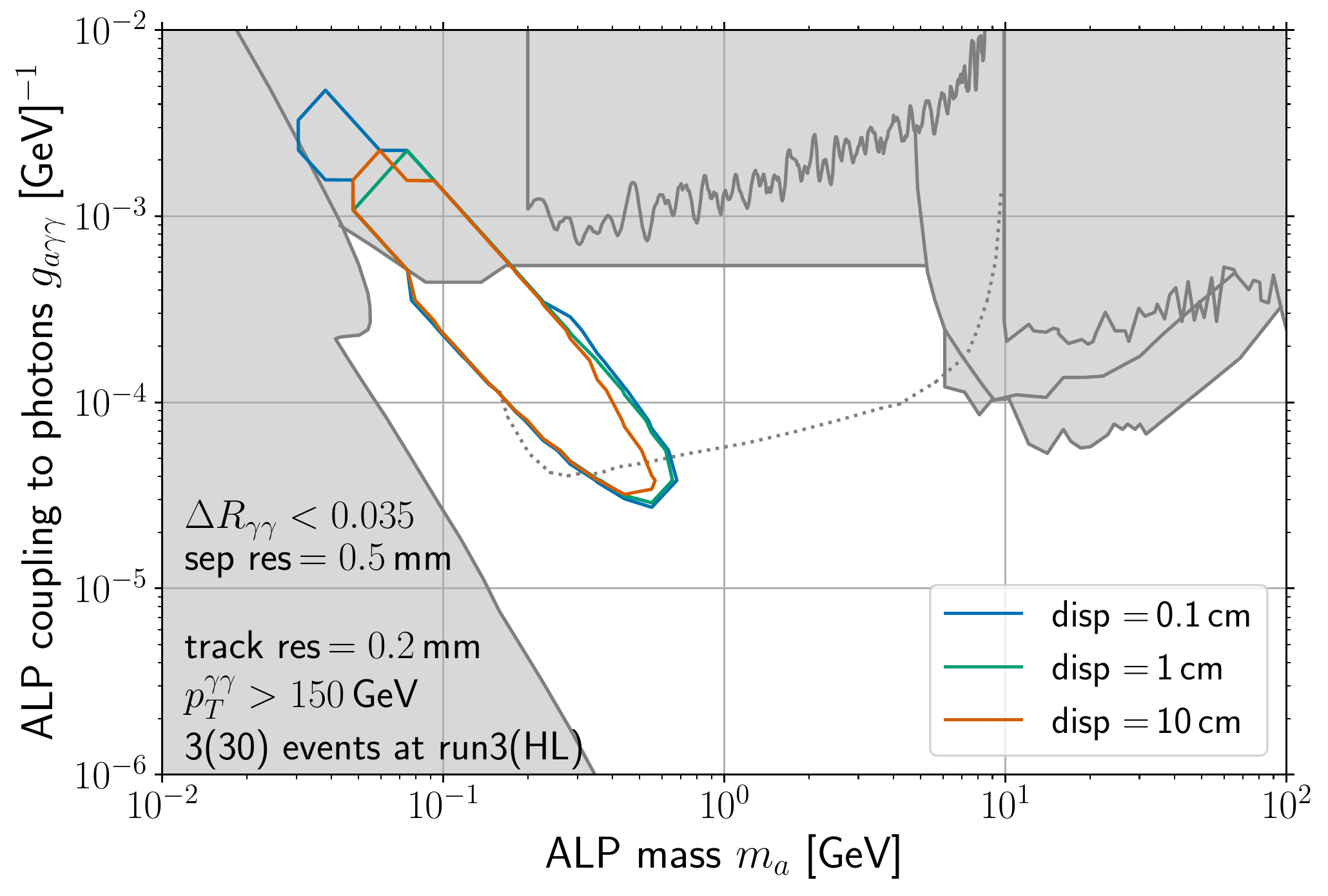}
\\
		\label{fig:vary_Delta_R1}
		\includegraphics[width=0.49\textwidth]{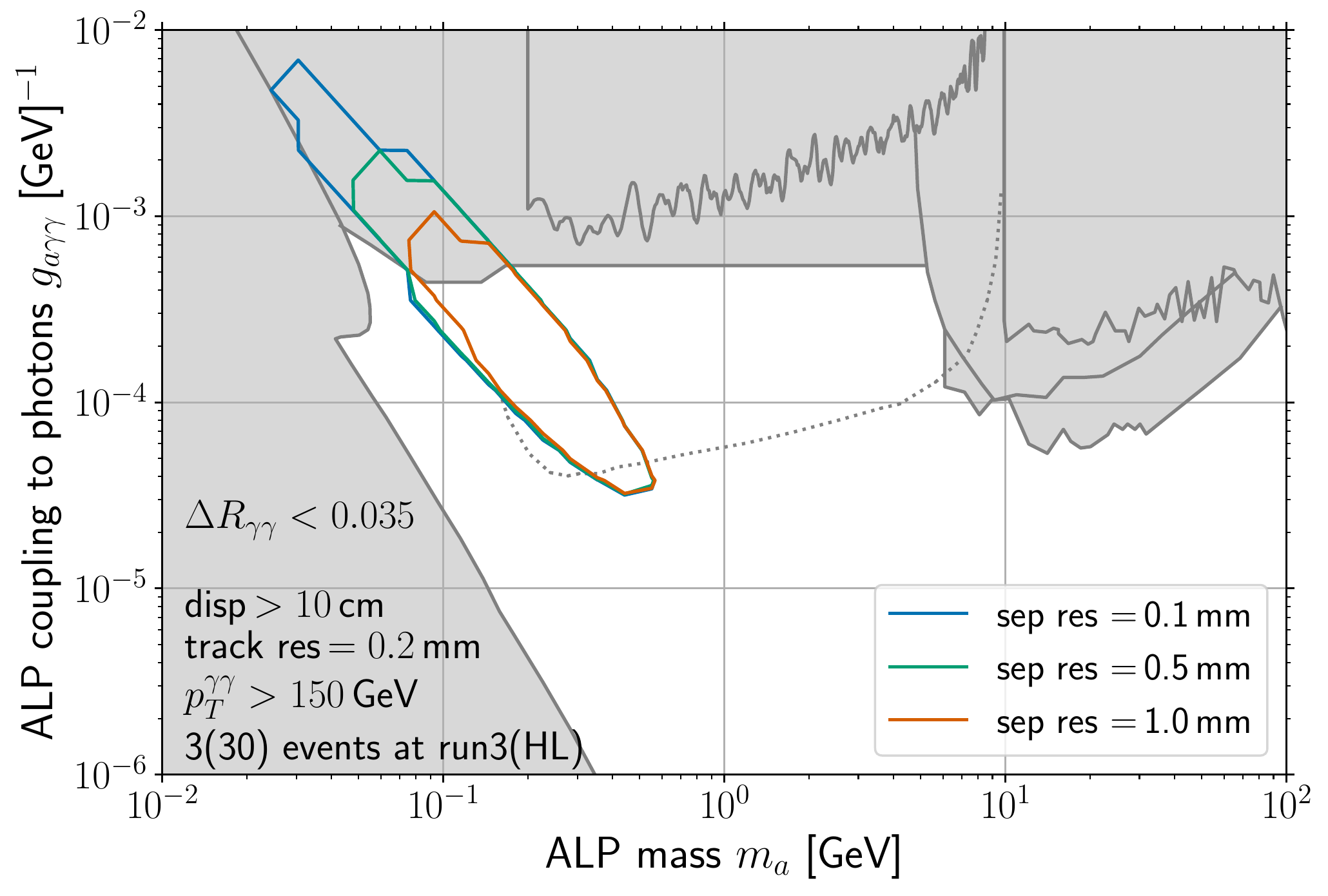}
&
		\label{fig:vary_Delta_R2}
		\includegraphics[width=0.49\textwidth]{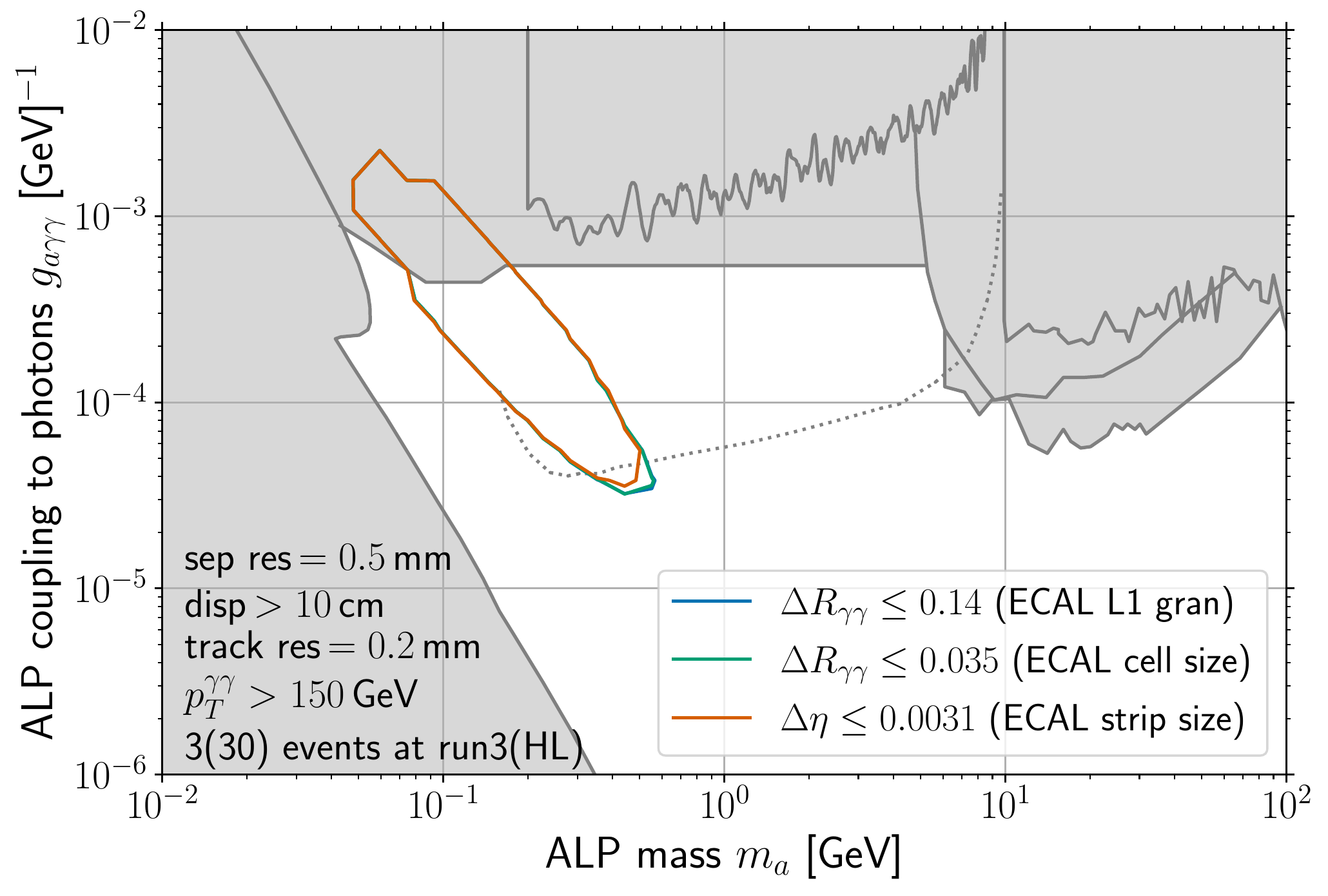}
\\
\end{tabular}
\vspace{-0.2cm}
\caption{Effect of varying different specifications on the reach of our proposed search strategy in the ALP mass vs.~photon coupling plane. From left to right and top to bottom, we examine the effects of changing the required number of events, the cut on the $p_T$ of the photon pair, the resolution of the ATLAS TRT, the minimum discernible vertex displacement, and the maximum angular separation of the individual photons within the pair.}
\label{fig:parameter_space_variations}
\end{figure}

In this section our goal is to assess which region of the mass vs.~photon coupling ALP parameter space can be probed using the search technique explained in the previous sections.
To do that, we simulate $10^6$ ALP productions and decays for each point in a grid of $(m_a, g_{a\gamma\gamma})$ values, and calculate the fraction of events that yield photon pairs that could be separately distinguishable at the tracker.
In the following we briefly describe the influence of the cuts and experimental parameters that we use in our analysis, and discuss how the sensitivity changes as they are varied.
Figure~\ref{fig:parameter_space_variations} shows how the reach of the search depends on the requirements set for each variable.
In each panel, we show the effect of varying a single quantity while keeping the rest of them fixed to a benchmark value.

\paragraph{Number of signal events.}
Given that we expect our proposed search to be background free, we follow the usual convention of requiring $3$ signal events for our sensitivity estimate.
This delineates the sensitivity reach shown in the upper left panel of Fig.~\ref{fig:parameter_space_variations}, where the orange line corresponds to the LHC Run-2 luminosity, and the green and blue lines to the Run-3 and HL-LHC ones, respectively.
That panel also gives an idea of how the sensitivity of the search decreases if we require more signal events, which could be necessary of additional sources of background were discovered.

\paragraph{Diphoton transverse momentum.}
A minimum transverse momentum of the photon pair is required for trigger purposes and $K_L$ background rejection.
We fix our benchmark value to $150$~GeV, which is slightly above the standard $140$~GeV ATLAS trigger on isolated photons during Run 2~\cite{ATLAS:2019dpa}.
Incidentally, this value results in a complete rejection of the $K_L\rightarrow\gamma\gamma$ background at the Run 2 luminosity (see Fig.~\ref{fig:KL_background}, right panel), ensuring a background-free search.
We stress, however, that our $K_L\rightarrow\gamma\gamma$ background estimate is based on an extrapolation of existing simulations of the $K_L$ angular and momentum spectrum.

The top left panel of Fig.~\ref{fig:parameter_space_variations} shows that the reach of the search increases slightly when the $p_T$ cut is lowered, as long as we assume the search to be background free.
We expect this to be a good approximation for any $p_T$ cut above $\sim 100$~GeV, but for lower values the $K_L$ background contamination will start reducing the reach of the search.
Determining the lowest $p_T$ cut that still allows for complete background rejection requires a dedicated, improved simulation of the $K_L$ which is out of the scope of this paper.

\paragraph{Track resolution.}
The spatial resolution with which the direction of a track can be determined is crucial for the displaced vertex determination, as explained in Sec.~\ref{sec:displaced}.
Being able to tag the ALP decays as displaced allows to reject the large background coming from promptly decaying hadronic states like $\pi^0\rightarrow\gamma\gamma$.
The nominal spatial resolution of the ATLAS TRT is $\sim 0.1$~mm~\cite{ATLAS:2017jwu,2017NIMPA.845..257M}. To be conservative we employ twice this value, i.e. $0.2$~mm, as our benchmark for the displaced vertex determination.
The dependence of our results on this resolution is showcased in the middle left panel of Fig.~\ref{fig:parameter_space_variations}.
A poorer spatial resolution significantly diminishes the reach of the search strategy, while assuming that the nominal resolution of $0.1$~mm can be achieved yields some modest gains.

\paragraph{Vertex displacement requirement.}
Closely related to the previously discussed quantity is the minimum distance from the primary interaction vertex to the ALP decay vertex that is required to tag the decay as displaced.
Although displaced-vertex searches at ATLAS have achieved sub-cm accuracy~\cite{ATLAS:2017tny}, to be conservative we set the minimum displacement to $10$~cm for our benchmark projections.
Then, on the middle right panel of Fig.~\ref{fig:parameter_space_variations} we show that allowing for smaller discernible displacements slightly increases the reach of the search towards heavier, more strongly coupled ALPs which have shorter decay lengths.

\paragraph{Separation resolution.}
It is important to understand how our results depend on the minimum separation between the photon tracks that can be distinguished.
We gauge this by placing a lower cut on the distance between the two collimated photons at the edge of the tracker.
To be conservative, we fix our benchmark value for the minimum separation to be $0.5$~mm, which is $5$ times the nominal TRT resolution of $0.1$~mm.
In the middle right panel of Fig.~\ref{fig:parameter_space_variations} we study how our results depend upon this value.
As expected, achieving the nominal resolution of $0.1$~mm allows the search to be extended to lighter ALPs whose decay products are more collimated.
Conversely, a poorer resolution to separate tracks at the TRT degrades the performance of our search strategy at small ALP masses.

\paragraph{Maximum angular separation.}
To make sure that the signal events are not rejected by the trigger requirements on the shower shape in the ECAL, we place a cut on the maximum angular separation $\Delta R_{\gamma\gamma}$ between the photons in the pair.
For our benchmark scenario we require that the photons satisy $\Delta R_{\gamma\gamma} \leq 0.035$, which corresponds to the angular size of a single ECAL cell in the second layer.
As can be seen in the right panel of Fig.~\ref{fig:parameter_space_variations}, loosening this requirement to $\Delta R_{\gamma\gamma} \leq 0.14$, the size of size of the ECAL towers used for the L1 trigger, does not appreciable affect the reach of the search strategy.
This would be enough for the event to pass the selection procedure at the hardware level and be recorded so that it is available for the modified reconstruction and identification process proposed in this paper.
As a matter of fact, tightening this requirement to $\Delta\eta\leq 0.0031$, the finest angular resolution that can be achieved at the ECAL, does not degrade the reach of the search.
The reason for this is that the vast majority of signal events have an angular separation that is well below this threshold.

\paragraph{Benchmark sensitivity scenarios.}

\begin{figure}[t!]
\centering
\includegraphics[width=0.8\textwidth]{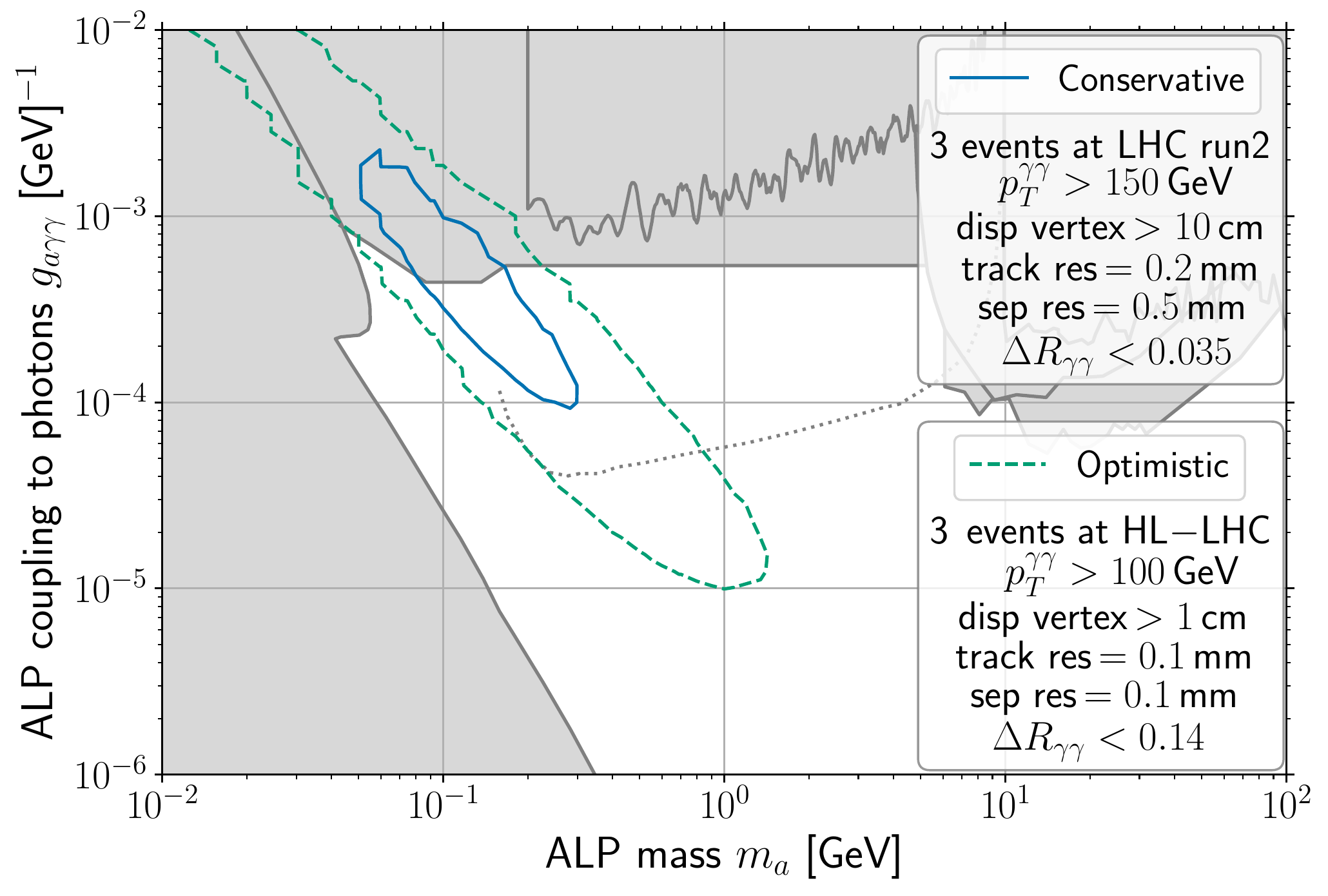} 
\caption{Potential reach of our proposed search strategy in the ALP mass vs. photon coupling parameter space. The solid line corresponds to a conservative estimate making use of existing data, while the dashed line is a more futuristic goal with more data, lower $p_T$ cuts, and highest resolution to distinguish displaced vertices.
Existing constraints are shaded in grey (see Fig.~\ref{fig:parameter_space_sketch}), and the grey dotted line is a projected limit using the full Belle-II luminosity~\cite{Dolan:2017osp}.}
\label{fig:conservative_optimistic}
\end{figure}

Combining the results of the paragraphs above, we construct two benchmark scenarios for the sensitivity of our proposed search strategy to test the ALP parameter space.
These correspond to a more conservative and a more optimistic set of choices for all the parameters discussed above, and their projected sensitivities are shown in Fig.~\ref{fig:conservative_optimistic}.
For our conservative estimate, we use the integrated luminosity of the Run-2 of the LHC, that is data that is already recorded and could be reanalyzed at present.
In contrast, our more optimistic forecast makes use of the full expected data to be collected by the high luminosity phase of the LHC.

Both benchmark scenarios showcase significant sensitivity to new parameter space, untested by other experiments and observations.
Interestingly and importantly, our search strategy is most sensitive in the region of relatively large couplings and masses, corresponding to relatively short decay lengths that are hard to access in fixed target experiments.
Our proposal therefore provides powerful complementarity, and offers the possibility to probe new parameter space using already collected data.

\section{Discussion and Conclusions}\label{sec:conclusions}

Axion-like particles in the mass range $10\,{\rm MeV}-10\,{\rm GeV}$ with moderately large photon couplings $g_{a\gamma\gamma}\sim (10^{-6}-10^{-3})\,{\rm GeV}^{-1}$ provide an interesting challenge for experimental searches.
This region of parameter space falls right between the traditional sensitivities of collider and fixed-target searches.
In this paper, we have investigated a method that may allow colliders, and in particular the LHC, to gain sensitivity in this region.
Our proposal overcomes the difficulties that LHC detectors have to separate highly collimated photon pairs coming from the decay of boosted (pseudo)scalar resonances.
Already existing data may allow to enter new, untested parameter space and future data from the HL-LHC will further increase the reach of the proposed search strategy.

The main idea is to use the exquisite position resolution of the tracker, whose angular resolution is two orders of magnitude better than that of the electromagnetic calorimeter commonly used for photon identification and isolation.
This greatly enhances the power to separate highly collimated photons, at the cost of requiring the photons to convert into $e^+e^-$ pairs before leaving the tracking detector.
We have shown that this strategy allows to separate the decay products of ALPs with masses as low as $10$~MeV.

Furthermore, the use of tracking information allows for a significantly better ability to discriminate displaced decays of particles into photons compared to using only calorimeter data.
This allows to overcome the main limitation for collider searches for particles with small displacements, and in our case is crucial to reject the background coming from promptly decaying hadronic resonances like $\pi^0$, which can also produce highly collimated photon pairs.
Our estimates, based on an extrapolation of existing data from~\cite{Kling:2021fwx}, suggest that irreducible backgrounds from long-lived neutral mesons such as $K_{L}$ are small enough in the target signal region, although further studies are required to make a final quantitative statement.
Overall, the analysis that we propose in this work can yield a significant gain in sensitivity and probe vast areas of untested ALP parameter space as shown in Fig.~\ref{fig:conservative_optimistic}.

The search strategy proposed in this work is not only applicable to ALPs, but can more generally be used to look for particles  decaying into photons with a small but macroscopic decay length. 
Thus, the development of methods based on tracking information can extend the range of sensitivity of LHC (and other colliders) to moderately long-lived particles decaying into photons.

It must be mentioned that our analysis is a rather crude theoretical study. To put this strategy into practice requires a closer investigation of potential experimental pitfalls. Importantly, we have assumed that events that are triggered due to sufficiently high transverse photon momenta and fulfilling simple photon isolation criteria are also successfully identified as photon candidates. 
This is rather non-trivial in the case of two photon events where both photons are converted inside the tracker. It is important to check whether these events that feature two pairs of tracks from the conversion per cluster in the electromagnetic calorimeter are not discarded or misidentified. Yet, as we have not made use of the tracker information for the trigger requirements we hope that, if necessary, a suitable modification of the analysis can be done even for the existing data. 
Finally, we note that a more careful experiment-level determination of (in principle) reducible backgrounds from misreconstruction and misidentification of particles is desirable.

Beyond locating displaced vertices, photons converted in the tracker may allow for an additional improvement. Due to the magnetic field, the conversion electrons and positrons curve, thus enabling a momentum measurement of the individual photons and, potentially, allowing for the reconstruction of the invariant mass of the decaying resonance. This could yield additional information on the origin of the two photons and a potential handle to further suppress backgrounds.

\section*{Acknowledgments}
We thank Monica Dunford for very useful discussions about the inner workings of the ATLAS detector, and Jim Cline for pointing out long-lived kaons as a background.
We are grateful to Felix Kling for providing us with the $K_L$ spectrum used to estimate this background. We would also like to thank Michael Spannowsky for optimistic comments about the possibility of doing an invariant mass reconstruction from the conversion tracks.
GA is supported by NSERC (Natural Sciences and Engineering Research Council, Canada), and gratefully acknowledges its efforts to facilitate parenthood at the postdoctoral stage through its parental leave scheme.
JJ would like to thank the EU for supporting the ITN HIDDeN, Marie Sklodowska-Curie grant agreement No 860881-HIDDeN.
DDL acknowledges the financial support from the Coordena\c{c}\~{a}o de Aperfei\c{c}oamento de Pessoal de N\'{i}vel Superior - Brasil (CAPES) - 23 Finance Code 001.

\bibliographystyle{utphys}
\bibliography{references}

\end{document}